\documentclass[preprint,10pt]{elsarticle}

\usepackage{soul}
\usepackage{graphicx}
\usepackage{amssymb}
\usepackage{amsthm}
\usepackage{amsmath}
\usepackage{booktabs}
\usepackage{graphicx}
\usepackage{float}
\usepackage{amsmath} 
\usepackage{multirow}
\usepackage{lineno}
\usepackage{array}
\usepackage{indentfirst}
\usepackage{supertabular}
\usepackage[colorlinks,
            linkcolor=red,
            anchorcolor=blue,
            citecolor=green
            ]{hyperref}
\usepackage{subfig}

\usepackage{algorithm}
\usepackage{algorithmicx}
\usepackage{algpseudocode}
\usepackage{amsmath}
\usepackage{amssymb}
\usepackage{bm}
\usepackage[table,x11names]{xcolor}
\usepackage{geometry}
\usepackage{graphicx}

\usepackage[nameinlink]{cleveref}
\crefname{figure}{Fig.}{Figs.}
\crefformat{equation}{Equqation ~#2(#1)#3}
\crefformat{section}{Section~#2#1#3}
\AtBeginDocument{%
	\let\citet\cite
}

\begin{document}

\begin{frontmatter}


\title{Multi-scale modeling in thermal conductivity of Polyurethane incorporated with Phase Change Materials using Physics-Informed Neural Networks}




\author[UMU,WM]{Bokai Liu}
\author[TH,WM]{Yizheng Wang}
\author[WM]{Timon Rabczuk}
\author[UMU]{Thomas Olofsson}


\author[UMU]{Weizhuo Lu\corref{spb}}

\cortext[spb]{Corresponding Authors; E-mail address: weizhuo.lu@umu.se}

\address[UMU]{Intelligent Human-Buildings Interactions Lab, Department of Applied Physics and Electronics, Umeå University, 901 87 Umeå, Sweden}

\address[WM]{Institute of Structural Mechanics, Bauhaus-Universit\"{a}t Weimar, Marienstr. 15, D-99423 Weimar, Germany}

\address[TH]{Department of Engineering Mechanics, Tsinghua University, Beijing, 100084, China}





\begin{abstract}

Polyurethane (PU) possesses excellent thermal properties, making it an ideal material for thermal insulation. Incorporating Phase Change Materials (PCMs) capsules into Polyurethane (PU) has proven to be an effective strategy for enhancing building envelopes. This innovative design substantially enhances indoor thermal stability and minimizes fluctuations in indoor air temperature. To investigate the thermal conductivity of the PU-PCM foam composite, we propose a hierarchical multi-scale model utilizing Physics-Informed Neural Networks (PINNs). This model allows accurate prediction and analysis of the material's thermal conductivity at both the meso-scale and macro-scale. By leveraging the integration of physics-based knowledge and data-driven learning offered by PINNs, we effectively tackle inverse problems and address complex multi-scale phenomena. Furthermore, the obtained thermal conductivity data facilitates the optimization of material design. To fully consider the occupants' thermal comfort within a building envelope, we conduct a case study evaluating the performance of this optimized material in a single room. Simultaneously, we predict the energy consumption associated with this scenario. All outcomes demonstrate the promising nature of this design, enabling passive building energy design and significantly improving occupants' comfort. The successful development of this PINNs-based multi-scale model holds immense potential for advancing our understanding of PU-PCM's thermal properties. It can contribute to the design and optimization of materials for various practical applications, including thermal energy storage systems and insulation design in advanced building envelopes.

\end{abstract}


\end{frontmatter}



\section{Introduction}
\label{S:1}

The rise in global energy consumption has sparked concerns about potential supply shortages, depletion of energy resources, and significant environmental consequences such as ozone layer depletion, global warming, and climate change \cite{olofsson2012modeling, olofsson2009building}. Alarming data collected by the International Energy Agency indicates that greenhouse gas emissions will continue to escalate due to global warming, resulting in severe weather patterns worldwide. To address these challenges, the European Union (EU) has set ambitious goals to reduce carbon footprints by 80\% to 95\% below 1990 levels by 2050 \cite{allard2013methods}. A key strategy in achieving a competitive low-carbon economy by 2050 is the improvement of energy performance in the building sector, which accounts for 36\% of total CO2 emissions. Notably, HVAC systems contribute to 50\% of the final energy consumption in the EU. Consequently, enhancing energy efficiency in this sector has gained significant attention as a popular research area \cite{sjogren2007approach}.

In recent decades, researchers have shown a growing interest in utilizing phase change materials (PCMs), specifically thermal energy storage, to enhance the energy efficiency of buildings \cite{zhou2021explicit, souayfane2016phase}. Incorporating PCMs into building materials or the building envelope has the potential to increase the capacity for latent heat storage, thereby improving the indoor thermal environment and comfort as well as energy efficiency\cite{michel2017experimental}. PCMs offer several advantages over other materials used for thermal energy storage, including a high heat of fusion, high energy storage density, and a constant phase change temperature \cite{tyagi2007pcm, pomianowski2013review}.

However, PCMs also face certain drawbacks, such as non-ideal thermal conductivity and the possibility of leakage during phase transitions \cite{yang2015rigid}. These limitations can be addressed by developing shape-stable polymer composites that include PCM inclusions and embedding them in building components to enhance the performance of the building envelope \cite{amin2014effective}. Microencapsulation, which involves encapsulating PCMs within polymeric shells, has emerged as a promising approach for confining PCMs in this application \cite{salunkhe2012review}. It satisfies the aforementioned requirements and offers additional benefits, such as affordability, low density, and low thermal conductivity of polymers. Among the available polymer matrices, polyurethane (PU) rigid foams have been widely employed for thermal insulation and are considered as excellent energy-saving materials \cite{nandy2023thermal}. The honeycomb-like structure of PU foams traps air, resulting in passive insulation properties and enhancing the heat absorption capacity of polyurethanes \cite{chen2021polyurethane}. PU-PCMs exhibit several advantages, including the lowest thermal conductivity (ranging between 0.02 and 0.05 W/mK), high mechanical and chemical stability at extreme temperatures, and the ability to form sandwich structures with various facer materials \cite{ikutegbe2020application}. In comparison to other insulating materials, PU-PCMs demonstrate high competitiveness.

Loucas Georgiou et al. examine the application of various PCM coatings in different environmental conditions and structural elements after construction. To determine the best conditions for PCM coatings, a validated numerical model based on finite elements and Life Cycle Assessment practices are employed holistically \cite{georgiou2023numerical}. U. R. Mahajan et al. develop a composite foam by incorporating microencapsulated n-tetradecane, a phase change material (PCM), into a polyurethane (PU) foam formulation. The PCM is microencapsulated with poly(methyl methacrylate-co-methacrylic acid) using oil-in-water emulsion polymerization \cite{mahajan2023development}.  Qingyi Liu et al. design self-healed inorganic phase change materials for thermal energy harvesting and management with experimental test \cite{liu2023self}. 
Francesco Galvagnini et al. develop polymeric insulating foams for low-temperature thermal energy storage applications, where different amounts of a microencapsulated paraffin having a melting temperature of -10 degree are used as phase change material (PCM) for the preparation of multifunctional rigid polyurethane foams (PUF) for refrigeration applications \cite{galvagnini2022development}. 

While previous studies have primarily focused on synthesis methods and experimental investigations of the thermal energy storage capacity of PU-PCM foams, there has been limited exploration of their thermal evaluation across multiple scales, which hinders a comprehensive understanding of the material's behavior and its complex system. To address this gap, this study aims to employ a multi-scale modeling approach to facilitate comprehensive research.

However, traditional multi-scale modeling methods often incur significant computational costs due to the requirement of numerous models at different scales, which limits the structural design of materials in complex multi-scale scenarios \cite{liu2023demat, liu2022surrogate}. Deep learning, a powerful method in machine learning, has achieved remarkable success across numerous domains. It has proven its effectiveness in computer vision \citep{ALEXNET}, speech recognition \citep{speech_recognition}, natural language processing \citep{machine_translation}, and strategic games \citep{alphago,star_game}, as well as in drug development \citep{alphafold}. Today, deep learning is ubiquitous, driving advancements and empowering various fields.
and improving energy efficiency is no exception \cite{pham2020predicting}. Physics-Informed
Neural Network (PINNs) and Deep Energy Method (DEM) \cite{samaniego2020energy, PINN_original_paper} utilize AI techniques to incorporate physical laws and data-driven learning in the combination with AI and science \cite{wang2022cenn, sun2023binn}.  Deep learning based on PINNs can use a data-driven approach combined with physical constitutive equations to simplify computational modeling and reduce costs while maintaining high accuracy. Since the introduction of physical laws, the algorithmic process is transparent and interpretable, which enhances the interpretability of traditional 'black box' machine learning.

The primary objective of this study is to develop a multi-scale model based on Physics-Informed Neural Networks (PINNs) that can accurately predict the thermal conductivity of PU-PCM (Polyurethane Phase Change Material). PINNs offer the advantage of integrating physics-based knowledge of the system with data-driven learning, making them well-suited for solving inverse problems \cite{lu2021physics} and handling complex multi-scale phenomena.
The multi-scale model aims to capture the intricate relationship between the microstructure and thermal properties of PU-PCM. It will utilize PINNs to incorporate the fundamental physics and governing equations governing heat transfer in the material. By leveraging available experimental data and simulations, the model will be trained to learn the relationship between the microstructural features and the resulting thermal conductivity.
The accuracy of the PINNs-based multi-scale model is crucial in understanding and predicting the thermal behavior of PU-PCM at different scales. The model's ability to handle noisy or limited data, as well as its capacity to provide uncertainty quantification, will be essential in dealing with real-world scenarios and ensuring reliable predictions.
The successful development of this PINNs-based multi-scale model has the potential to advance our understanding of PU-PCM's thermal properties and contribute to the design and optimization of materials for various practical applications, such as thermal energy storage systems, building insulation, and electronic cooling.

Based on the above situation, this paper focuses on the PINNs-based multi-scale modeling of PU-PCM at different scales, and conducts a case study to implement the impact of this composite on the energy efficiency of the building envelope.  This article is structured as follows. \Cref{sec: 2} presents the general methodology, followed by an introduction of the multi-scale modeling including PINNs enhanced and RVE-FEM method in Section \Cref{sec: 3}. Engineering application for a case study is illustrated in \Cref{sec: 4}, and the numerical results are presented in Section \Cref{sec: 5}. Finally, the manuscript concludes with Section \Cref{sec: 6}, which outlines the conclusions.

\section{Methodology of research}
\label{sec: 2}

\begin{figure}[htp]
	\centering\includegraphics[width=0.6\linewidth]{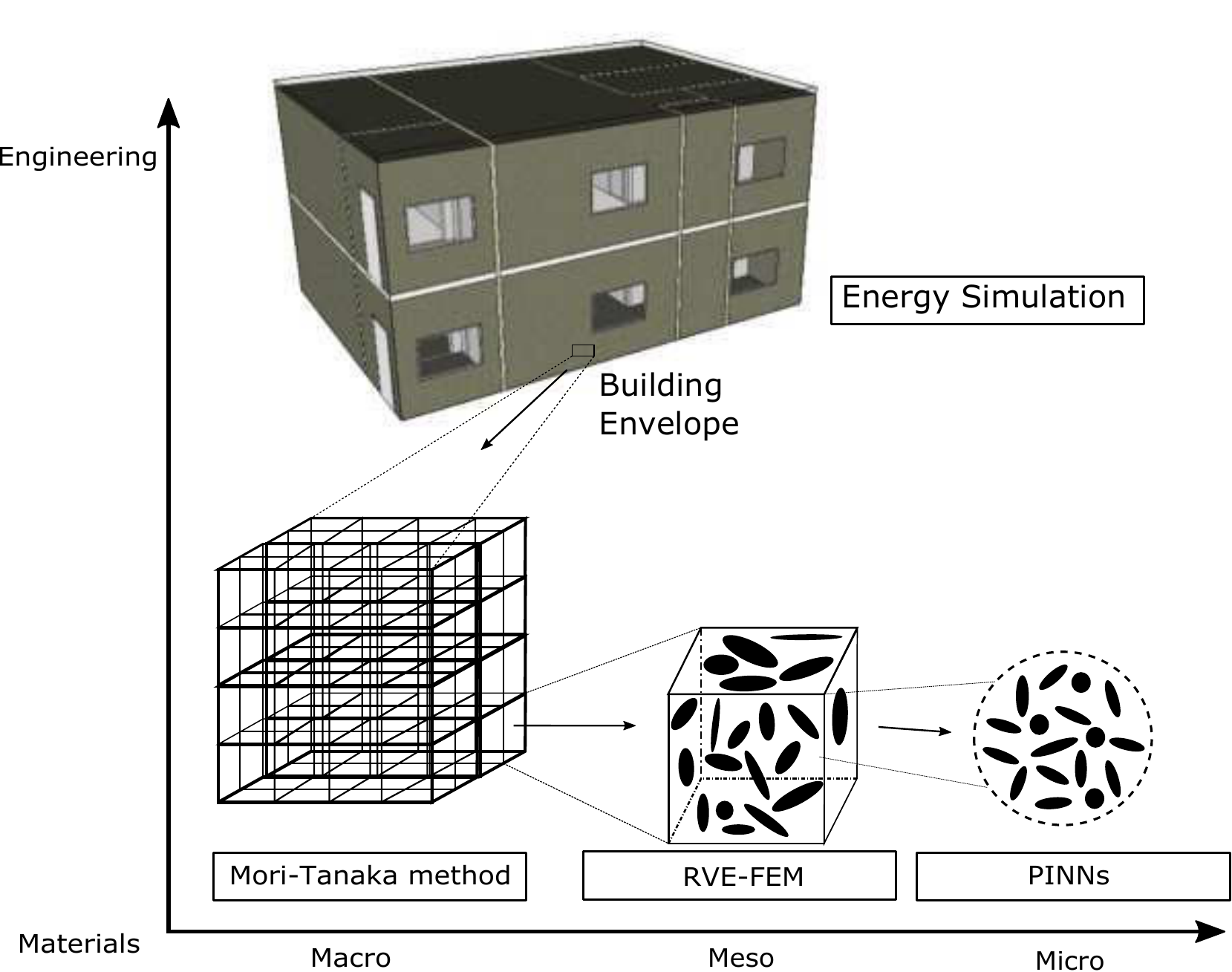}
	\caption{Multi-scale modeling scheme}
	\label{fig: Multiscaleflowchart}
\end{figure}

We propose a multi-scale modeling approach for studying the behavior of the building envelope, incorporating Physics-Informed Neural Networks (PINNs) and a RVE-FEM (Representative Volume Element - Finite Element Method) method. This modeling strategy employs a hierarchical approach, bridging three different length scales from the micro to macro scales, as illustrated in \Cref{fig: Multiscaleflowchart}. Our approach begins with a bottom-up approach that transfers information through the length scales \cite{liu2022stochasticfull} . In this hierarchical framework, the output of the finer scale serves as the input for the next coarser scale. Finally, we employ an engineering application to analyze the output from the multi-scale model. Our approach involves four main steps, as outlined below:

1) \textbf{Bottom-up modeling}: We start with a bottom-up approach, where we consider the finest scale of the system and gather information about its behavior. This step helps us understand the micro-scale characteristics and phenomena.

2) \textbf{Micro-scale modeling using PINNs}: We employ Physics-Informed Neural Networks (PINNs) to model the micro-scale behavior. PINNs allow us to capture the underlying physics of the system and incorporate it into the neural network architecture. This step helps us simulate and analyze the behavior at a micro-scale level.

3) \textbf{Meso \& Macro-scales modeling using RVE-FEM}: We then move on to the meso and macro scales, where we consider larger representative volumes of the system. The RVE-FEM method is used to model the behavior of these larger volumes, taking into account the information obtained from the micro-scale modeling. This step allows us to bridge the gap between micro and macro scales and capture the overall behavior of the system.

4) \textbf{Engineering application for a Building Envelope}: Finally, we apply our multi-scale model to analyze the behavior of the building envelope. This includes assessing its thermal properties, energy consumption, and other relevant engineering parameters. By incorporating the results obtained from the micro and meso-macro scale modeling, we can provide insights into the performance of the building envelope and optimize its design and functionality.

This multi-scale approach enables a comprehensive understanding of the building envelope's behavior and provides valuable insights for engineering applications.

\section{Multi-scale modeling}
\label{sec: 3}

In this study,  we consider composites materials design by introducing PU (polyurethane) as the matrix of the composite. At the same time, we seal paraffin PCM with polymer materials to prevent leakage during phase transition. Microencapsulation of PCMs can make it to be stable energy storage. This design can both reduce the thermal conductivity and seal the PCMs.

\begin{figure}[htp]
	\centering\includegraphics[width=1\linewidth]{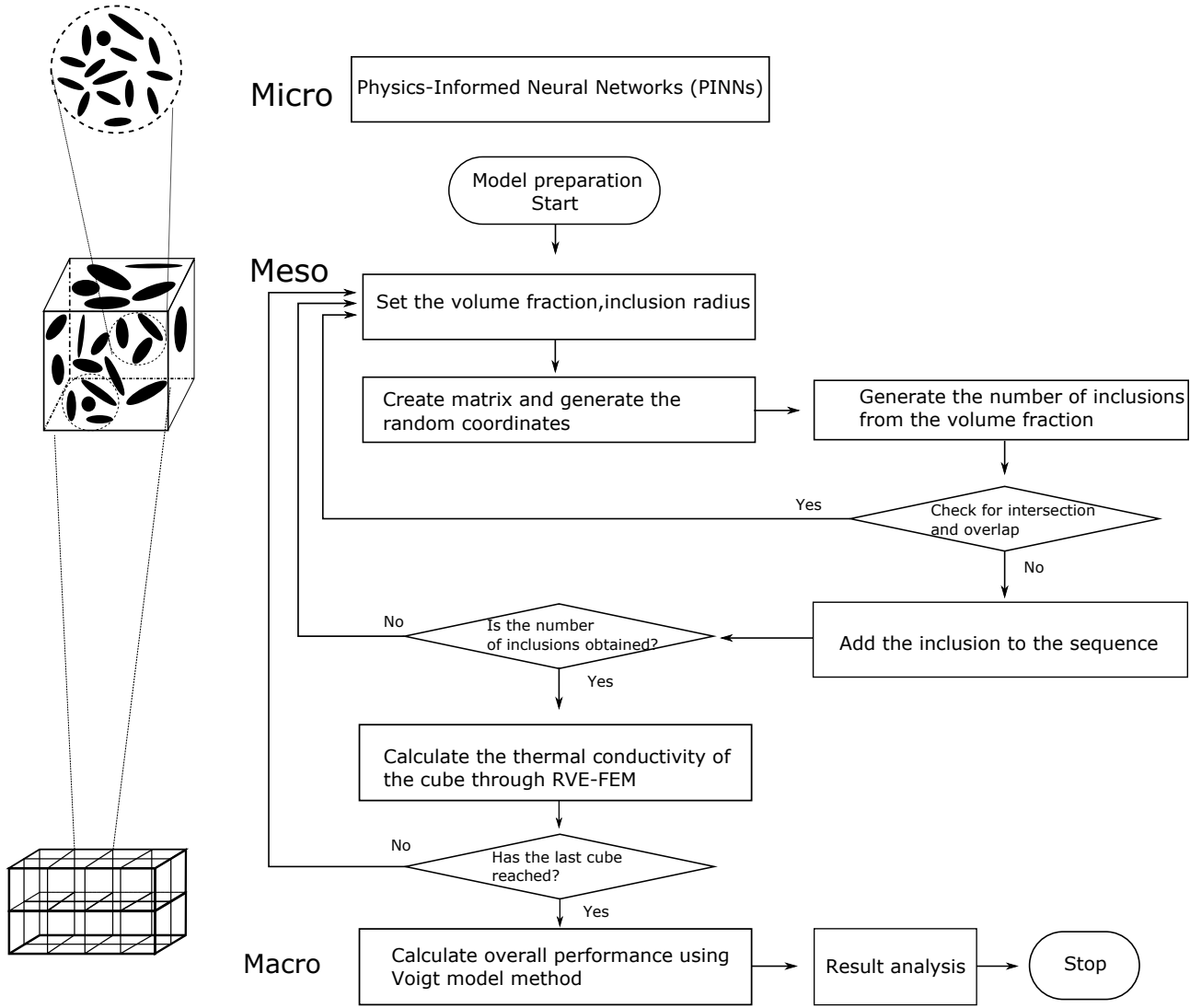}
	\caption{Multi-scale modeling scheme}
	\label{fig: MultiscaleMesoMacro}
\end{figure}

We adopt a bottom-up approach using a hierarchical multiscale method. This approach involves transferring information exclusively from the fine scale to the next coarser scale. The flowchart illustrating this process is depicted in \Cref{fig: MultiscaleMesoMacro}. We now proceed to describe the models employed at different length scales within our multiscale framework.

\subsection{ Physics-Informed Neural Networks (PINNs)-based Micro-scale modeling }
\label{S:3.1}
\begin{figure}
	\begin{centering}
		\subfloat{\centering{}\includegraphics{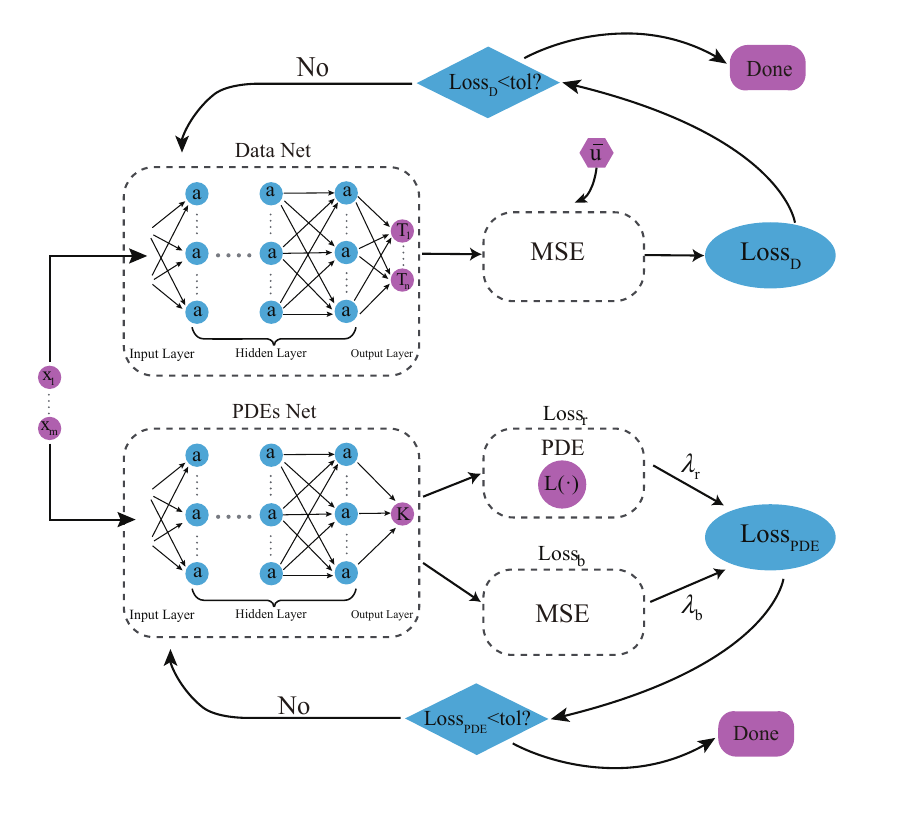}}
		\par\end{centering}
	\caption{Schematic of Physics-Informed Neural Networks (PINNs)-based Micro-scale modeling, the left purple circles  in the neural network are the inputs (temporal and spatial coordinates).
		The right blue circles  in the neural network are the hidden neurons. The right  purple circles of the Data Net
		in the neural network are the output (interesting field). $\boldsymbol{L(\cdot)}$ is the differential
		operator related to the PDEs. $\bar{\boldsymbol{u}}$ is the given data. MSE is the mean square error.
		$\lambda_{r}$ and $\lambda_{b}$ are the weight of the residual loss $Loss_r$
		and the boundary loss $Loss_b$ (including initial condition if temporal problem)
		respectively. Data Net and PDEs Net updates the parameters of neural network until the loss is less than the threshold.
		\label{fig:Schematic-of-PINN}}
\end{figure}
In this section, the outline for Physics-Informed Neural Networks (PINNs)-based Micro-scale modeling is presented, as shown in  \Cref{fig:Schematic-of-PINN}. The method uses a feed-forward neural network for multi-scale modeling. The neural network is a multiple linear regression with the activation function to enhance non-linear capabilities. The feed-forward neural network is represented as:

\begin{equation}
\begin{split}\boldsymbol{z}^{(1)} & =\boldsymbol{w}^{(1)}\cdot \boldsymbol{x}+\boldsymbol{b}^{(1)}\\
\boldsymbol{a}^{(1)} & =\boldsymbol{\sigma}(\boldsymbol{z}^{(1)})\\
\vdots\\
\boldsymbol{z}^{(L)} & =\boldsymbol{w}^{(L)}\cdot \boldsymbol{a}^{(L-1)}+\boldsymbol{b}^{(L)}\\
\boldsymbol{a}^{(L)} & =\boldsymbol{\sigma}(\boldsymbol{z}^{(L)})\\
\boldsymbol{y} & =\boldsymbol{w}^{(L+1)}\cdot \boldsymbol{a}^{(L)}+\boldsymbol{b}^{(L+1)}
\end{split}
\end{equation}

Here, $\boldsymbol{z^{(l)}}$ is the linear transformation of the previous neurons $\boldsymbol{a^{(l-1)}}$, where layers $1\leq l\leq L$ are the hidden layers. The output $\boldsymbol{a^{(l)}}$ of $\boldsymbol{z^{(l)}}$ is obtained through the activation function $\sigma$, which is typically a non-linear function like tanh or sigmoid. In this study, tanh is used as the activation function:

\begin{equation}
tanh=\frac{e^{x}-e^{-x}}{e^{x}+e^{-x}}.
\end{equation}

The feed-forward neural network serves as an approximation function to fit experimental or observed data in the Data Net. The discrepancy between the data and neural network approximation, represented as $Loss_{D}$, is commonly evaluated using Mean Square Error (MSE) as the criterion. The construction of the loss function is guided by the specific Partial Differential Equations (PDEs) related to the problem. The loss function consists of a boundary loss $Loss_{b}$ and a domain loss $Loss_{r}$ representing the unknown conductivity field, approximated by the PDEs Net. The parameters of PDEs Net, approximating the unknown conductivity, are optimized using the PDEs.

In this specific case, the one-dimensional steady-state heat conduction equation with an unknown conductivity field $k(x)$ is considered:
\begin{equation}
\begin{cases}
k(x)\frac{d^{2}T}{dx^{2}}=f(x) & x\in[a,b]\\
-\frac{\partial T}{\partial x}=q_{1} & x=a\\
\frac{\partial T}{\partial x}=q_{2} & x=b
\end{cases}.
\end{equation}
The objective is to develop Physics-Informed Neural Networks (PINNs) to model and predict $k(x)$ with  the given temperature distribution $T(x)$ in the domain $[a, b]$, boundary condiction ($q_{1}$ and $q_{2}$), and heat source term $f(x)$.

To achieve this, two neural networks are constructed:
\begin{itemize}
	\item \textbf{Data net $\hat{T}(x_{i};\boldsymbol{\theta}_{D})$}: This neural network approximates the temperature distribution $T(x)$ at given spatial coordinates $x_i$. The parameters of the Data Net are denoted as $\boldsymbol{\theta}_{D}$. The data $\{x_{i},T_{i}\}_{i=1}^{n}$, where $T_i$ is the observed temperature at position $x_i$, are used to train the Data Net. The data loss, represented as $Loss_{D}$, is calculated using Mean Square Error (MSE):
	\begin{equation}
	Loss_{D}=\sum_{i=1}^{n}||\hat{T}(x_{i};\boldsymbol{\theta}_{D})-T(x_{i})||^{2}.
	\end{equation}
	\item \textbf{PDEs Net $k(x_{i};\boldsymbol{\theta}_{k})$}: This neural network approximates the unknown conductivity field $k(x)$. The parameters of the PDEs Net are denoted as $\boldsymbol{\theta}_{k}$. The loss of the PDEs net, represented as $Loss_{PDE}$, is constructed based on the PDEs and boundary conditions:
	\begin{equation}
	Loss_{PDE}=\lambda_{1}\sum_{i=1}^{n}||\hat{T}(x_{i};\boldsymbol{\theta}_{D})-T(x_{i})||^{2} + \lambda_{2}\sum_{i=1}^{n}||k(x_{i};\boldsymbol{\theta}_{k})\frac{d^{2}\hat{T}(x_{i}\boldsymbol{\theta}_{D})}{dx^{2}}-f(x_{i})||^{2}+\lambda_{3}||\frac{d\hat{T}(a;\boldsymbol{\theta}_{D})}{dx}+q_{1}||^{2}+\lambda_{4}||\frac{d\hat{T}(b;\boldsymbol{\theta}_{D})}{dx}-q_{2}||^{2}.
	\end{equation}
\end{itemize}

The weight parameters $\lambda_{1}$, $\lambda_{2}$, $\lambda_{3}$, and $\lambda_{4}$ control the importance of each term in the loss function. Determining the appropriate weights for PINNs is crucial. In our approach, we employ an adaptive weight method based on the gradient. Specifically,

The weight of PINNs is an essential problem. As a result, we use the adaptive weight method according to different loss. we adjust the weight of PINNs according to gradient \citet{RN553}. To be specific, 
\begin{equation}
	\lambda_{i}=Factor*\frac{\sum_{j=1}^{4}mean\{|\nabla_{nn}Loss_{j}|\}}{mean\{|\nabla_{nn}Loss_{i}|\}},
\end{equation}
where Factor is the scale determined by the specific problem. In our method, the Factor is setted to 1.

The optimization algorithm is then used to find the optimal values of the parameters $\boldsymbol{\theta}_{D}$ for the Data Net and $\boldsymbol{\theta}_{k}$ for the PDEs Net. The objective is to minimize the losses $Loss_{D}$ and $Loss_{PDE}$:

\begin{equation}
\boldsymbol{\theta}_{D}=\arg\min_{\boldsymbol{\theta}_{D}}Loss_{D}(\boldsymbol{\theta}_{D};\{x_{i},T_{i}\}_{i=1}^{n})
\end{equation}

\begin{equation}
\boldsymbol{\theta}_{k}=\arg\min{\boldsymbol{\theta}_{k}}Loss_{PDE}(\boldsymbol{\theta}_{k};\{x_{i},f_{i}\}_{i=1}^{n},q_{1},q_{2},\boldsymbol{\theta}_{D}).
\end{equation}

Note that if the conductivity field $k(x)$ is a constant, it can be represented by a single optimization variable, denoted as $k$. There is no need to use a neural network to approximate it. The optimization algorithm would then involve finding the optimal value of this scalar variable $k$ to minimize the PDEs loss.

\subsection{Meso-scale modeling}
\label{S:3.2}

\begin{figure}[htbp]
	\centering
	\begin{minipage}[t]{0.48\textwidth}
		\centering
		\includegraphics[width=0.8\linewidth]{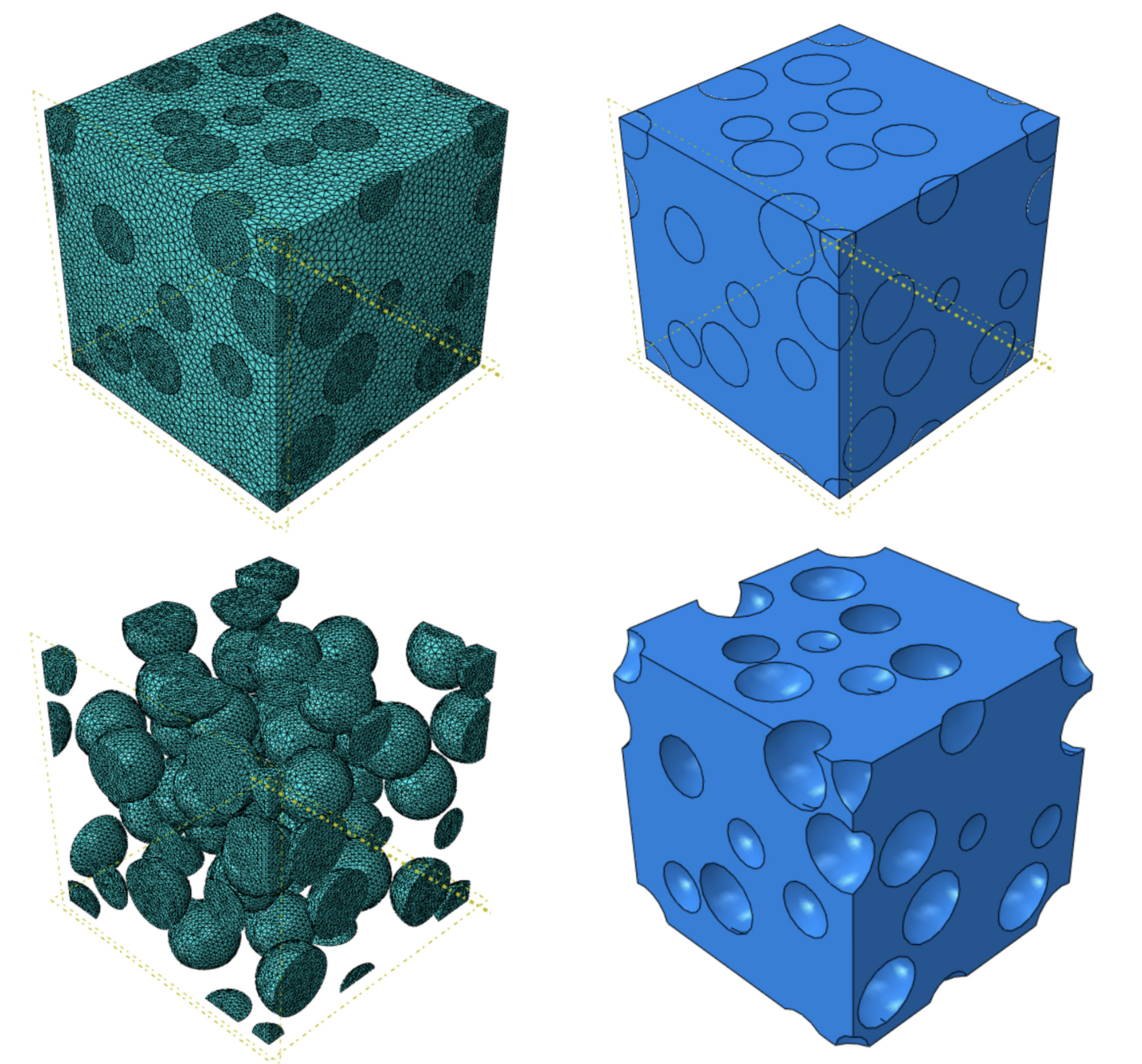}
		\caption{The RVE cube and Meshing}
		\label{fig: composite}
	\end{minipage}
	\begin{minipage}[t]{0.5\textwidth}
		\centering
		\includegraphics[width=6cm]{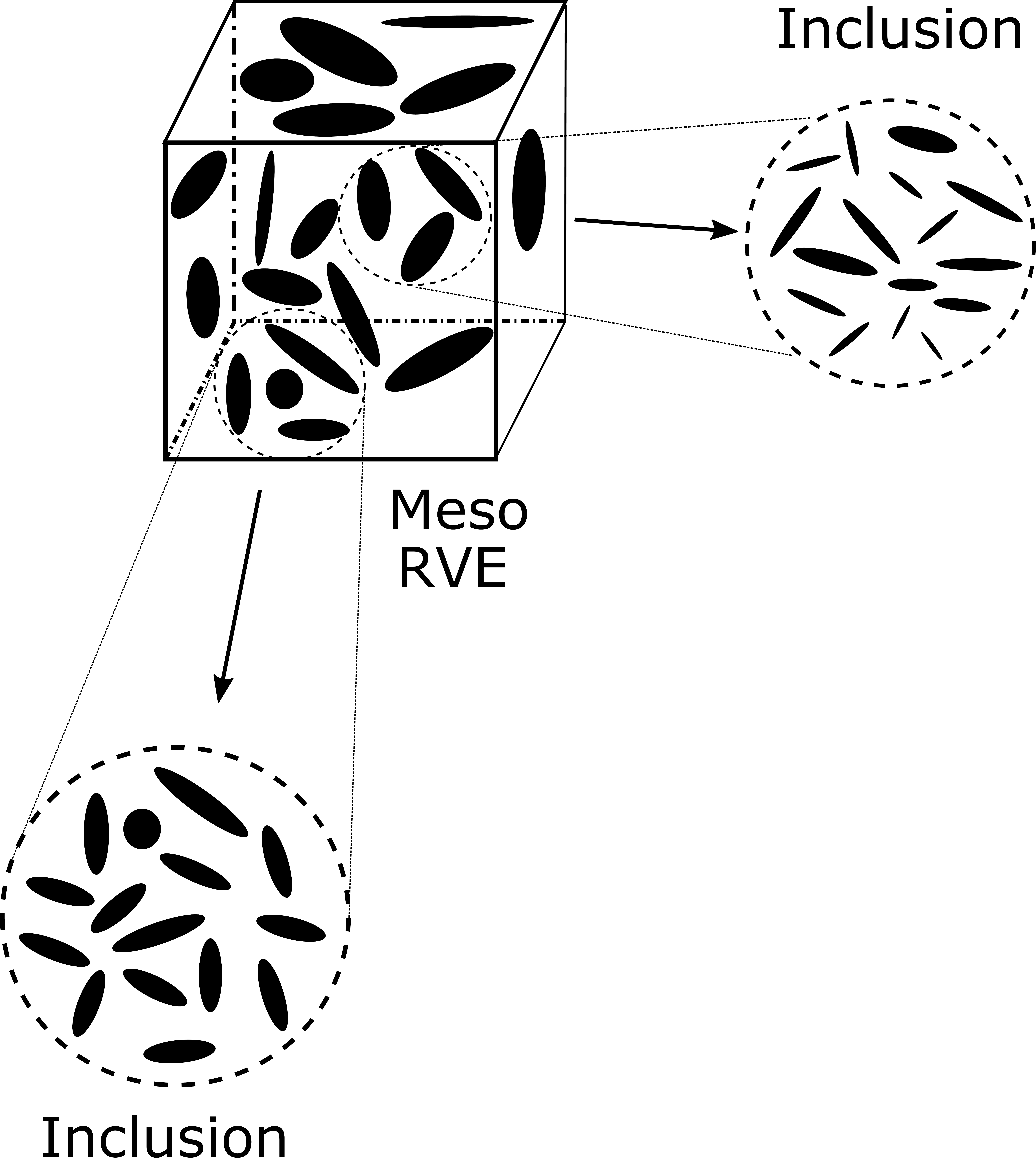}
		\caption{The agglomeration and dispersion in the Cubic RVE}
		\label{fig: Inclusion}
	\end{minipage}
\end{figure}

Our multi-scale modeling approach employs continuum models at the meso-scale, utilizing Representative Volume Elements (RVEs) that contain a limited number of inclusions capable of accurately representing the material properties. In our method, we commonly assume a cubic RVE, as shown in \Cref{fig: composite}, while simplifying the PCM fillers within the RVE as sphere. To facilitate the generation of RVE structures, we leverage the commercial software package Abaqus in conjunction with a custom Python script \cite{liu2020stochastic, liu2021stochastic, liu2022stochastic}. This script automates the creation of RVE structures by utilizing a 3D non-collision algorithm written in C++ \cite{mortazavi2013modeling, he2016modeling}.

The placement of fillers within the RVE is determined based on given probability density functions (PDFs) associated with the input parameters. This allows us to accurately incorporate the statistical distribution of the fillers. A visual representation of our meso-scale modeling approach is provided in Fig \Cref{fig: MultiscaleMesoMacro} at the top, illustrating the workflow. Additionally, the discretization of the model utilizes quadratic tetrahedral elements, as depicted in Fig \Cref{fig: composite}.

At the meso-scale, we take into account the presence of agglomerations and dispersions of PCMs spheres, which can occur when there is a high aspect ratio and high volume fraction. To analyze and quantify the degree of agglomeration, we employ a two-parameter method \cite{vu2015uncertainty}. This method involves generating distinct spheres that represent gathered zones, as depicted in Fig \Cref{fig: Inclusion}. The spherical regions within these zones are considered as the 'inclusions'. The entire space is divided into two components - $V_{Graphene}^{inclusion}$ and $V_{Graphene}^{matrix}$:

\begin{equation}
\label{eq:VolumeCNT}
V_{Graphene}=V_{Graphene}^{inclusion}+V_{Graphene}^{matrix}
\end{equation} 
where $V_{Graphene}^{inclusion}$ and $V_{Graphene}^{matrix}$ represent the volumes occupied by the spheres placed in the inclusions and matrix, respectively. We define the agglomeration index $\xi$ and dispersion index $\zeta$ as follows:
\begin{equation}
\label{eq:TwoparametersMeth}
\xi = \frac{V_{inclusion}}{V}, \qquad \zeta = \frac{V_{Graphene}^{inclusion}}{V_{Graphene}} 
\end{equation}  
The agglomeration index $\xi$ is defined as the volume fraction of inclusions relative to the total volume of the RVE. On the other hand, we utilize the dispersion index $\zeta$ to represent the inner volume fraction of PCM spheres within the inclusions, relative to the total volume of the composite. In the case where $\xi = \zeta$, it indicates that the spheres are uniformly distributed throughout the RVE, signifying the absence of agglomeration. Conversely, when $\xi > \zeta$, it suggests that the spheres are unevenly spaced within the RVE, indicating the presence of agglomeration.

\begin{figure}[htbp]
	\centering
	\begin{minipage}[t]{0.48\textwidth}
		\centering
		\includegraphics[width=8cm]{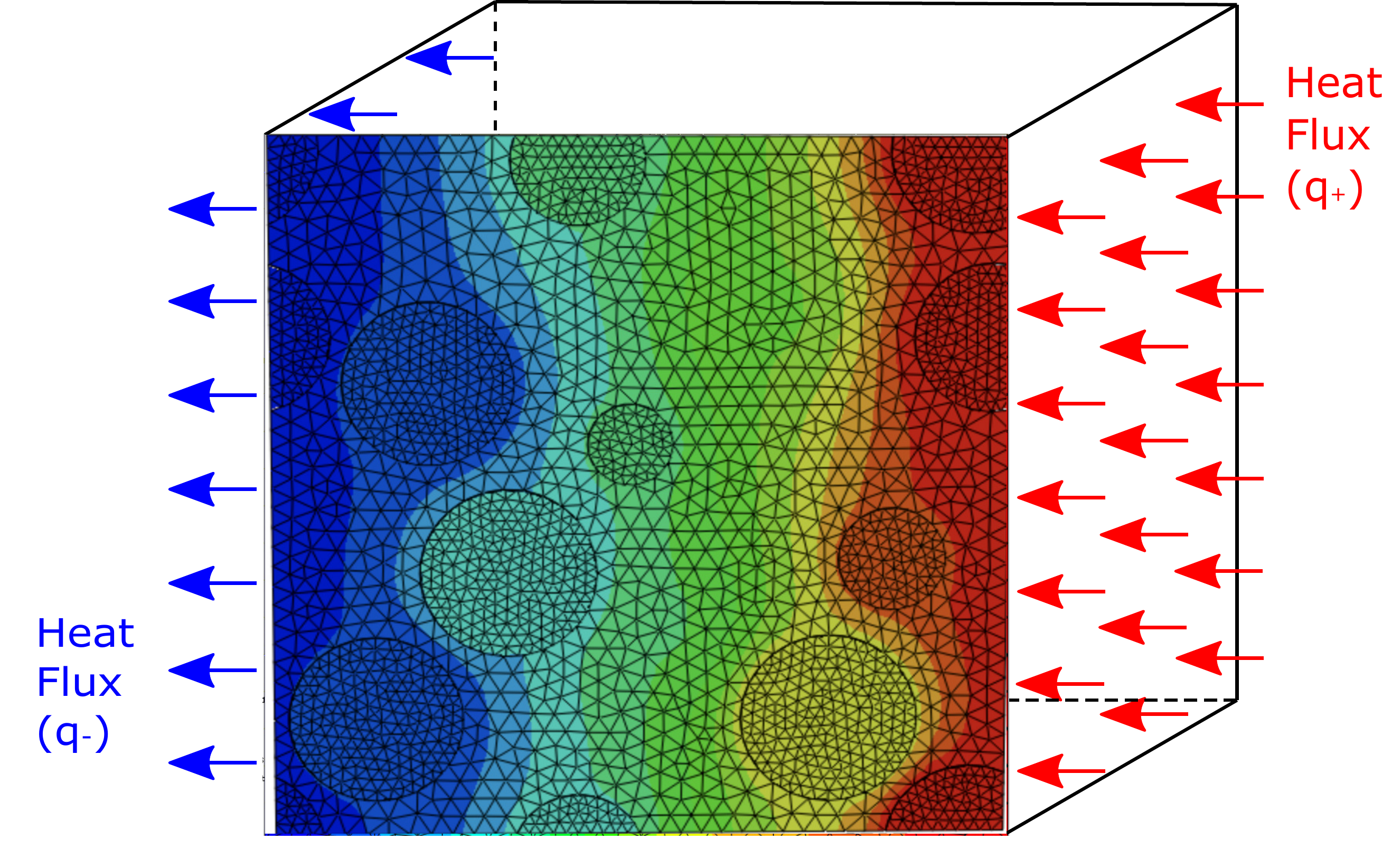}
		\caption{Applying heat flux in both sides}
		\label{fig: HotCold}
	\end{minipage}
	\begin{minipage}[t]{0.5\textwidth}
		\centering
		\includegraphics[width=6cm]{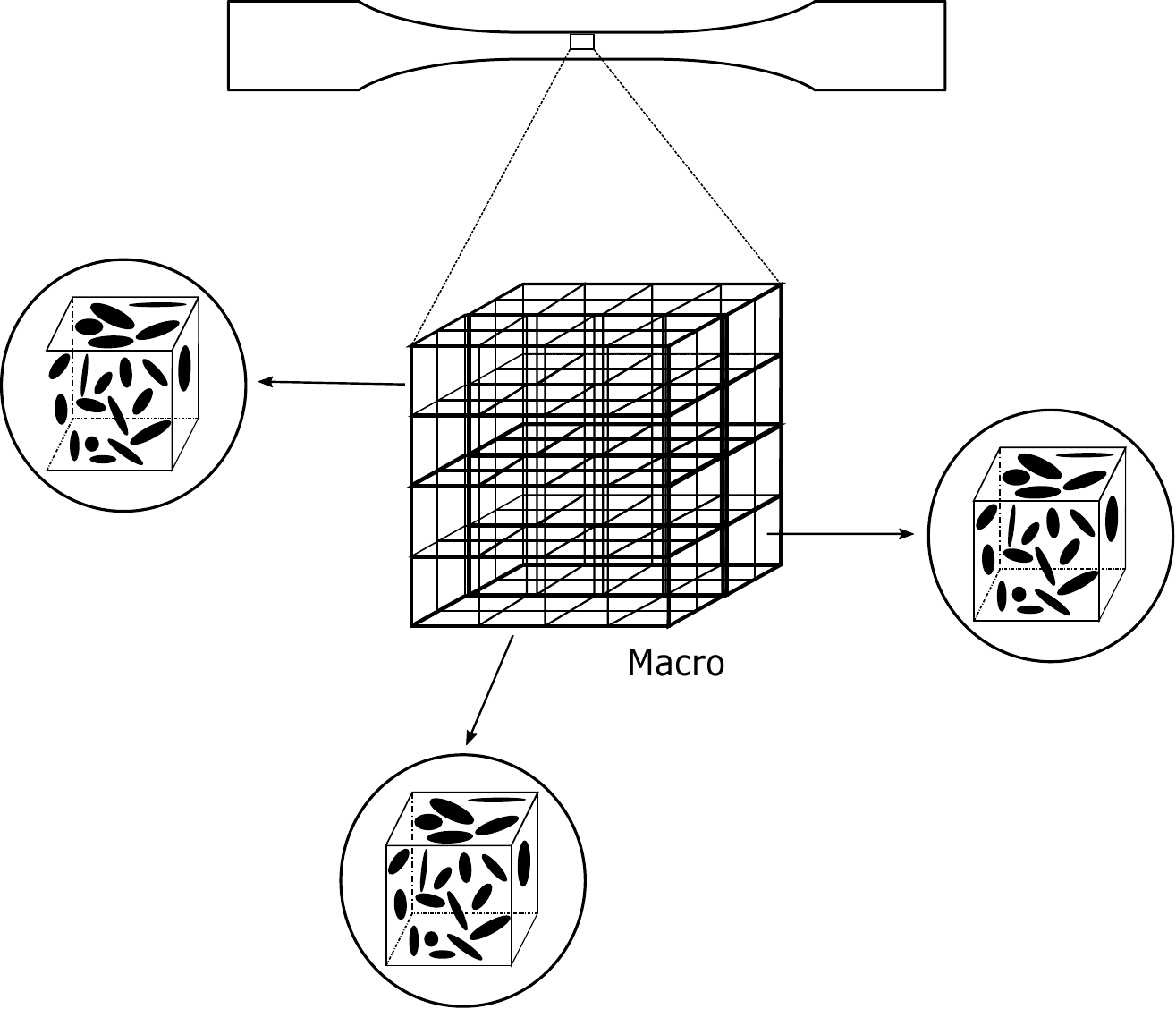}
		\caption{The material region in macro-scale modeling}
		\label{fig: MacroModel}
	\end{minipage}
\end{figure}

The initial step in the process is to determine an appropriate size for the Representative Volume Element (RVE). One approach to achieve this is by employing the sample enlargement method. This method entails gradually increasing the size of the RVE until the homogenized thermal conductivity reaches a desired level of convergence. The convergence criterion is determined by averaging the thermal conductivity values over a significant number of samples. By monitoring the homogenized thermal conductivity and observing its convergence to a specific value, we can establish a suitable RVE size for the analysis: 
\begin{equation}
\label{eq:<R>}
\langle R  \rangle= \frac{1}{M}\sum_{K=1}^{M}{R^{(K)}}
\end{equation}  \\
where $R^{(k)}$ is the current value in the $k-th$ RVE, and $M$ is the total RVE number. After the ensemble is averaged, a convergence criterion must be satisfied to define a suitable RVE size: 
\begin{equation}
\label{eq:<RConvergence>}
\bigg\lvert \frac{\langle R^{(K+1)}\rangle - \langle R^{(K)}\rangle}{\langle R^{(K)}\rangle} \bigg\rvert  < Tol = 1\%
\end{equation}  \\
where $R^{(k)}$ is the current value in the $k-th$ RVE, and  $R^{(k+1)}$ is the $k+1-th$ RVE. The heat transfer problem is governed by 
\begin{equation}
\label{eq:local equation}
C_f \frac{\partial\theta}{\partial t}+ \nabla\cdot \boldsymbol{q}- Q=0  
\end{equation} 
where $\theta$ represents the absolute temperature and $Q$ is the heat source. The heat capacity is denoted by $C_f$, and the heat flux vector is given by $\boldsymbol{q}$. For quasi-steady problems, the time-dependent term $C_f \frac{\partial\theta}{\partial t}$ is commonly neglected. Substituting Fourier's law into the governing equation yields
\begin{equation}
\label{eq:governing}
div (\kappa \nabla \theta) + Q = 0  \qquad in\quad\Omega
\end{equation}
with natural boundary conditions
\begin{equation}
\label{eq:PBCs q}
q_n=-\boldsymbol q\boldsymbol\cdot\boldsymbol n={\overline q}   \qquad on \quad \Gamma_q
\end{equation}
where $\boldsymbol{n}$ is the normal vector and ${\overline q}$ is the flux at boundary ${\Gamma_q}$. The  weak form of the heat equation is given by: Find  $\theta \in \nu$  such that:

\begin{equation}
\label{eq:wekform}
{\int_{\Omega}} \kappa \nabla \theta \cdot \nabla\delta \theta  d\Omega =
-{\int_{\Gamma_q}} \delta \theta\bar{q} d\Gamma
+ \int_{\Omega} \delta \theta Q d\Omega \quad  \forall \delta\theta \in \nu_0
\end{equation}
with $\theta \in \nu$ and $\delta \theta \in \nu_0$, in which $\theta$ denotes the trial function and $\delta\theta$ the test function. As \Cref{fig: HotCold} suggests, we apply two different heat fluxes from one direction through the cubic RVE results in a temperature gradient. We then apply Fourier's law to compute the homogenized thermal conductivity:
\begin{equation}
\label{eq:constitutive equation}
\boldsymbol{q}=-\boldsymbol{\kappa} \nabla T
\end{equation}
\begin{equation}
\label{eq:K}
\mathrm{with}\quad\boldsymbol{\kappa}=\begin{Bmatrix}\kappa_{xx}&0&0\\0&\kappa_{yy}&0\\0&0&\kappa_{zz}\end{Bmatrix}
\end{equation}  
where $\kappa_{xx}=\kappa_{yy}=\kappa_{zz}$ is defined by applying boundary conditions at different RVE edges and $\kappa$ is the conductivity of the composite. The macroscopic thermal conductivity of the composite is the output at the meso-scale.

\subsection{Macro-scale modeling}
\label{S:3.1.2} 

The Mori-Tanaka method is a widely adopted model for predicting the effective thermal conductivity of composite materials \cite{mortazavi2013modeling}. It offers a framework to estimate the overall thermal conductivity of a composite by considering the properties of its constituents and their spatial arrangement.  At the macroscopic level, a larger structure is considered to encompass uncertainties, and it is homogenized by randomly distributing cubes with distinct thermal properties obtained from mesoscale simulations, as illustrated in \Cref{fig: MacroModel}.

In the context of thermal conductivity, the Mori-Tanaka method takes into account the thermal conductivities of the matrix material and the reinforcement material, along with their respective volume fractions within the composite. The method employs an averaging technique to estimate the effective thermal conductivity, assuming that the composite's thermal conductivity is governed by two primary mechanisms: the thermal conductivity of the matrix and the improved thermal conduction pathways facilitated by the reinforcement.

The equation for estimating the effective thermal conductivity of a PU-PCMs enhanced composite using the Mori-Tanaka method is as follows:

\begin{equation}
\label{eq:MoriTanaka}
k_{\text{eff}} = k_m + \frac{{4k_pV_p}}{{(1-V_p) + \frac{{k_m}}{{k_p}}}}
\end{equation}
where $k_\text{eff}$ is the effective thermal conductivity of the composite; $k_m$ is the thermal conductivity of the polymer matrix; $k_p$ is the thermal conductivity of paraffin, and $V_p$ is the volume fraction of PCMs in the composite.

\section{PU-PCMs application: Case study}
\label{sec: 4}

\begin{figure}[htbp]
	\centering
	\begin{minipage}[t]{0.48\textwidth}
		\centering
		\includegraphics[width=8cm]{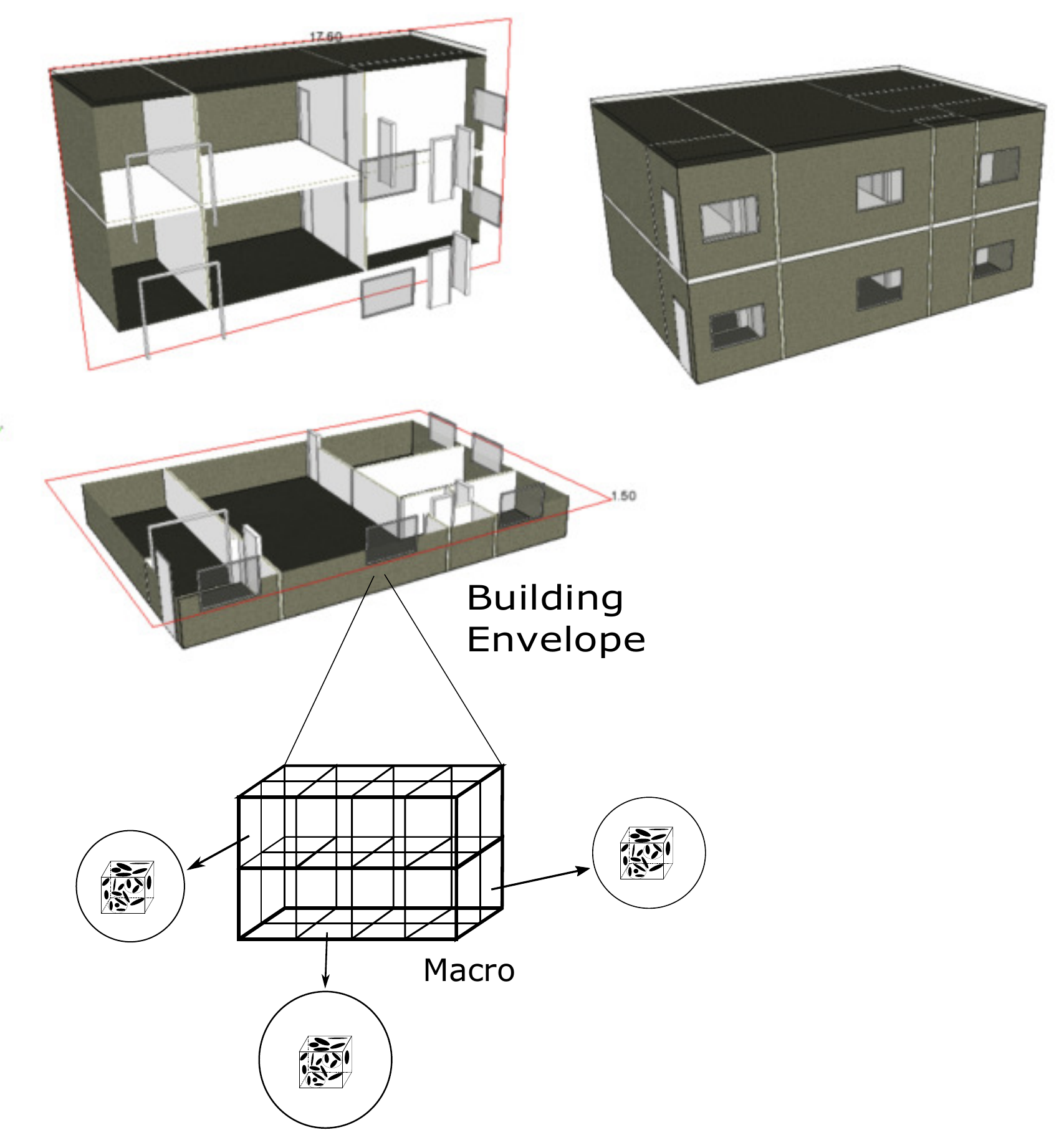}
		\caption{The application of PU-PCMs in building envelope}
		\label{fig: EngApplication}
	\end{minipage}
	\begin{minipage}[t]{0.48\textwidth}
		\centering
		\includegraphics[width=8cm]{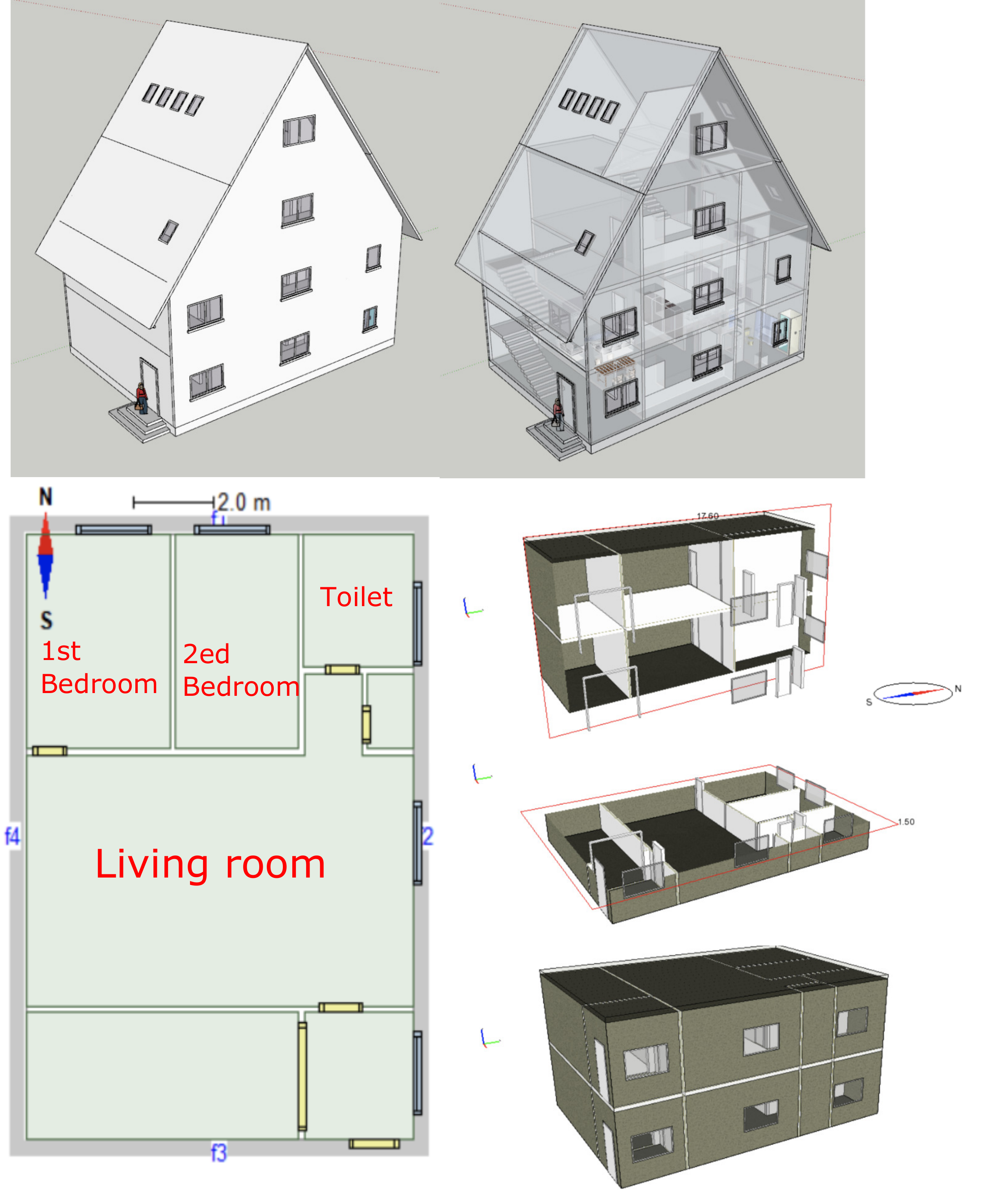}
		\caption{Architectural plans for singe-family house}
		\label{fig: BuildingPlan}
	\end{minipage}
\end{figure}

We have selected a single-family house that represents a common architectural style found in European regions. The application of PU-PCMs is illustrated in \Cref{fig: EngApplication}, where the composite is employed in the building envelope. The house is designed with two floors and features various rooms and spaces, as illustrated in \Cref{fig: BuildingPlan}. These include a living room, a first bedroom on the first floor, a second bedroom on the first floor, a stairwell, a bathroom, a storage room, a third bedroom on the second floor, and a fourth bedroom on the second floor.

\begin{figure}[htbp]
	\centering
		\begin{minipage}[t]{0.48\textwidth}
		\centering
		\includegraphics[width=6cm]{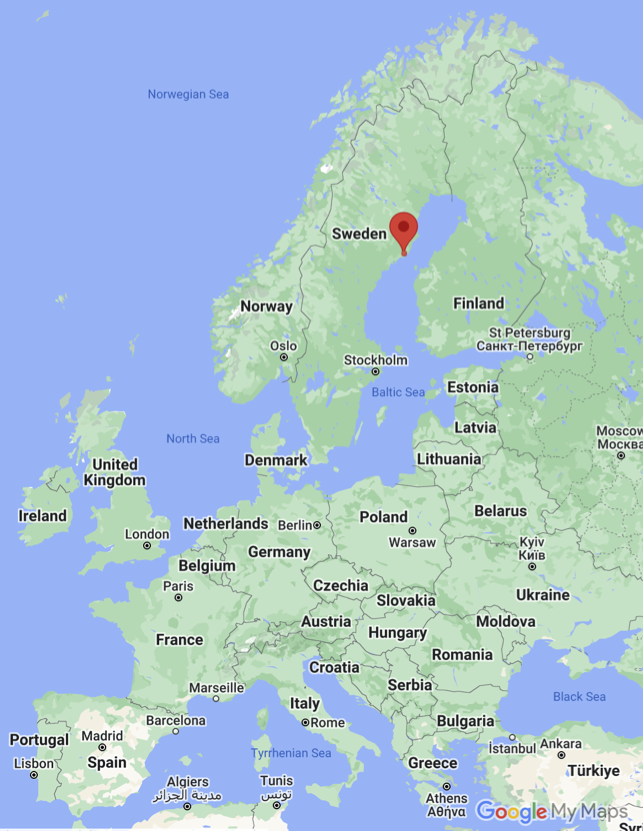}
		\caption{Location of Umeå, Västerbotten County, Sweden (Source: Google Maps)  }
		\label{fig: Location}
	\end{minipage}
	\begin{minipage}[t]{0.48\textwidth}
		\centering
		\includegraphics[width=9cm]{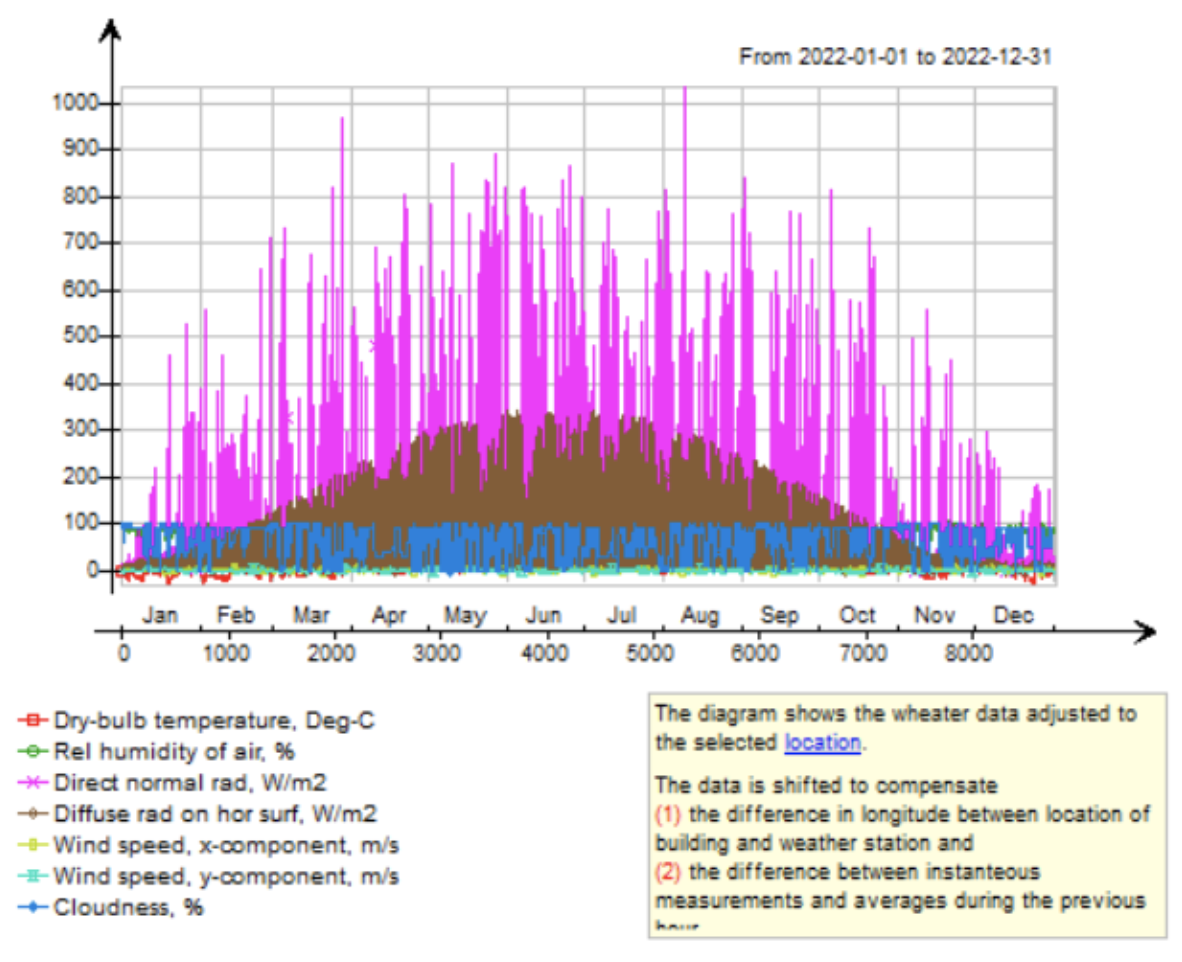}
		\caption{The climate distribution of Umeå in 2022}
		\label{fig: Climate}
	\end{minipage}

\end{figure}

The specific location of the house is in Umeå, Sweden, which is a typical city situated in the northeastern part of the country, shown in \Cref{fig: Location}. Umeå serves as the capital of Västerbotten County. The city experiences a subarctic climate classified as Dfc, which borders on a humid continental climate (Dfb). Umeå has relatively short yet moderately warm summers, while its winters are long and freezing, although typically milder compared to areas at the same latitude with a more continental climate.

The average temperature in January, which is the coldest month, is around -7$^\text{o}$C. In contrast, during July, which is the warmest month, the average temperature rises to about 16 $^\text{o}$C. The climate distribution of Umeå in the year of 2022 is shown in \Cref{fig: Climate}. In fact, the city has experienced exceptionally high temperatures during certain periods. The record high temperature of 32.2$^\text{o}$C was recorded on July 23, 2014, during an unusually warm summer in Sweden. Due to the climate change and global warming, the temperature is going up year by year, which means maintaining indoor thermal stability and comfort is highly demanded.

Based on the location information and climate conditions mentioned earlier, this case study focuses on the energy consumption for heating houses in the Umeå area. To address this, we explore the application of Phase Change Materials (PCMs) specifically Polyurethane Phase Change Materials (PU-PCMs). PU-PCMs have the ability to store and release thermal energy through phase change, while also reducing the U-value (thermal transmittance) of the house. In this scenario, we propose adding an additional layer of PU-PCM on the interior walls and ceiling of the house. The engineering parameters of PU-PCMs are derived from previous multiscale modeling approaches, ensuring their accurate representation.

Our simulation is primarily based on the hourly temperature trends observed in Umeå throughout the year 2022, considering it as the external temperature load. The variations in indoor temperature are predominantly influenced by the changes in external temperature as well as the heating provided by fuel sources. By analyzing these factors, we aim to evaluate the effectiveness of incorporating PU-PCMs in reducing energy consumption and enhancing thermal comfort within the house during the summer period.

When evaluating building performance in this case study, we consider several indicators, including building annual energy consumption, Predicted Percentage of Dissatisfied (PPD), Predicted Mean Vote (PMV), and annual acceptable comfort hours. Where the Fanger indices (PMV, PPD) are defined as follows \cite{fanger1967calculation}:

\begin{equation}
{\text{PMV}}=\left(0.303{e}^{-2.5128}+0.028\right)\left\{\begin{array}{c}366076.58368558-\\ 69.79815056{P}_{a}+0.09772{t}_{a}\\ -4.356\times {10}^{-8}\left[\begin{array}{c}{\left(0.198{t}_{r}+{t}_{a}+273\right)}^{4}\\ -{\left({t}_{r}+273\right)}^{4}\end{array}\right]\\ -1.17612{t}_{r}\end{array}\right\}
\end{equation}
where ${P}_{a}$is a function of relative humidity, partial pressure of water vapor in air, and $\varphi$ is relative humidity ${P}_{a}=\left(\varphi \right)$. Therefore, under certain conditions of indoor environment, PMV can be expressed as ${\text{PMV}}=\left(v,{T}_{g},{t}_{a},\varphi \right)$ \cite{li2023data}. Besides, PPD is calculated by the value of PMV, shown below:

\begin{equation}
{\text{PPD}} = 100 - 95*\exp \left( { - 0.03353*{\text{PMV}}^{4} - 0.2179*{\text{PMV}}^{2} } \right)
\end{equation}

Based on the ASHRAE 55 standard, PPD must be less than 20\% and PMV value tends to 0 can be considered thermally comfortable \cite{de2002thermal}.

\section{Numerical results and discussion}
\label{sec: 5}

\subsection{Multi-scale modeling results}
\label{S:5.1}

\subsubsection{Micro scale results based on  PINNs}
\label{S:5.1.1}
To demonstrate the outstanding performance of the PINNs-based micro-scale model in solving inverse problems, we consider two common scenarios: a uniform thermal conductivity scenario and a non-uniform thermal conductivity scenario. It is worth noting that the non-uniform scenario aligns with the characteristics of composite materials PU-PCMs where different materials have different thermal conductivities.

\begin{figure}
	\begin{centering}
		\subfloat{\centering{}\includegraphics{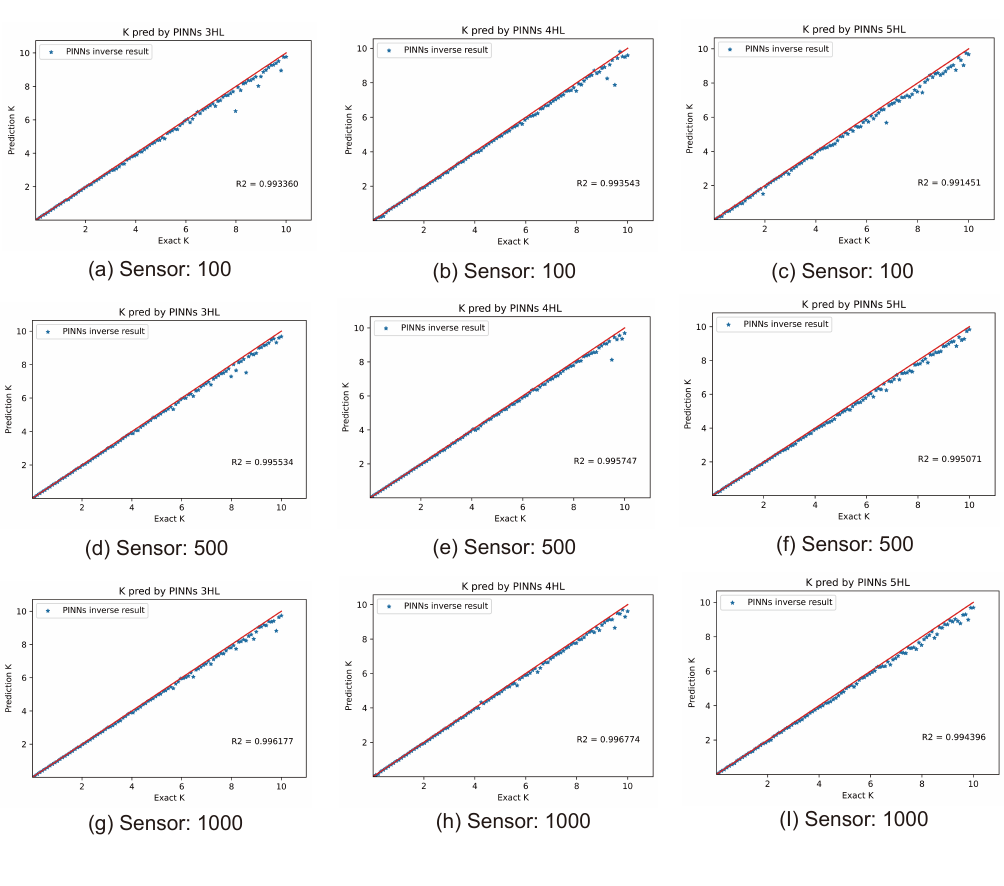}}
		\par\end{centering}
	\caption{The result of prediction for  uniform conductivity by  PINNs-based micro-scale model. The number of sensor is 100 (a, b, c), 500 (d, e, f), 1000 (g, h, i). The neural network with 3 hidden layers (a, d, g); The neural network with 4 hidden layers (b, e, h); The neural network with 5 hidden layers (c, f, i).
		\label{fig:k_constant}}
\end{figure}

For the uniform problem, the equation is as follows:

\begin{equation}
\begin{cases}
k\frac{d^{2}T}{dx^{2}}=f(x)=-(15\pi)^{2}cos(15\pi x) & x\in[-\frac{1}{2},\frac{1}{2}]\\
\frac{\partial T}{\partial x}=-q_{1}=-\frac{15\pi}{k} & x=-\frac{1}{2}\\
\frac{\partial T}{\partial x}=q_{2}=\frac{15\pi}{k} & x=\frac{1}{2}
\end{cases}.
\end{equation}
The analytical solution is given by 

\begin{equation}
T=\frac{cos(15\pi x)}{k}.
\end{equation}
We use the analytical solution to generate temperature field data. The objective is to predict the thermal conductivity $k$ using PINNs, given the temperature field $T(x)$, heat source term $f(x)$, and the boundary conditions $q_{1}(x=-1/2)$ and $q_{2}(x=1/2)$. The temperature field $T(x)$ and heat source term $f(x)$ are obtained from sensors uniformly distributed in the range [-1/2, 1/2]. The results as shown in \Cref{fig:k_constant} are shown for 100, 500, and 1000 sensor points, and we also test the influence of different neural network structures (3 hidden layers, 4 hidden layers, 5 hidden layers) on the results. The coefficient of determination $R^{2}$ demonstrates that PINNs achieve excellent results in predicting the uniform thermal conductivity ($R^{2}$ greater than 0.99). This is attributed to the powerful fitting ability of neural networks and the incorporation of physical principles in PINNs. It is worth noting that traditional multiscale micro-scale modeling algorithms for solving inverse problems involve significant computational effort \cite{fish2010multiscale}. In contrast, our approach using PINNs provides results for every thermal conductivity inverse problem in just a few minutes. Our results demonstrate the great potential of combining PINNs with multi-scale modeling for solving inverse problems in PU-PCMs.

In engineering practice, measured data from sensors generally contain noise due to systematic and experimental errors. To simulate the measured data in industrial applications, random noise is introduced to describe the experimental noise. For simplicity and generality, a zero-mean Gaussian distribution is applied to describe the signal noise, with the probability density given by:

\begin{equation}
\rho(x)=\frac{1}{\sqrt{2\pi\sigma_{s}}}exp(\frac{(x_{s}-\mu_{s})^{2}}{2\sigma_{s}}).
\end{equation}

Considering the real thermal conductivity of PCMs materials is 0.24, \Cref{tab:k_constant_noisy} show the results of the PINNs-based micro-scale model with different perturbations under $k=0.24$. \Cref{tab:k_constant_noisy} demostrates that even with significant perturbations, the PINNs-based micro-scale model still achieves high accuracy. Therefore, the PINNs-based micro-scale model is suitable for solving practical engineering problems.

\begin{table}
	\caption{The $L_{1}^{rel}$ error of uniform thermal conductivity with noisy data. $\sigma_{s}$ denotes the variance of Gaussian distribution of noisy data. HL denotes the different number of hidden layers.}\label{tab:k_constant_noisy}
	\begin{centering}
		\begin{tabular}{cccccccccccc}
			\toprule 
			$\sigma_{s}$ & 0 (no noise) & 0.01 & 0.02 & 0.03 & 0.04 & 0.05 & 0.06 & 0.07 & 0.08 & 0.09 & 0.10\tabularnewline
			\midrule
			3HL & 0.0288 & 0.0264 & 0.0402 & 0.0407 & 0.0523 & 0.069 & 0.0543 & 0.0471 & 0.0512 & 0.0588 & 0.0614\tabularnewline
			4HL & 0.0175 & 0.0353 & 0.0501 & 0.0378 & 0.0421 & 0.527 & 0.0544 & 0.0324 & 0.0296 & 0.0494 & 0.0661\tabularnewline
			5HL & 0.0297 & 0.0695 & 0.0345 & 0.0421 & 0.0531 & 0.0396 & 0.0579 & 0.0402 & 0.0747 & 0.0540 & 0.0683\tabularnewline
			\bottomrule
		\end{tabular}
		\par\end{centering}
	
\end{table}

\begin{figure}
	\begin{centering}
		\subfloat{\centering{}\includegraphics{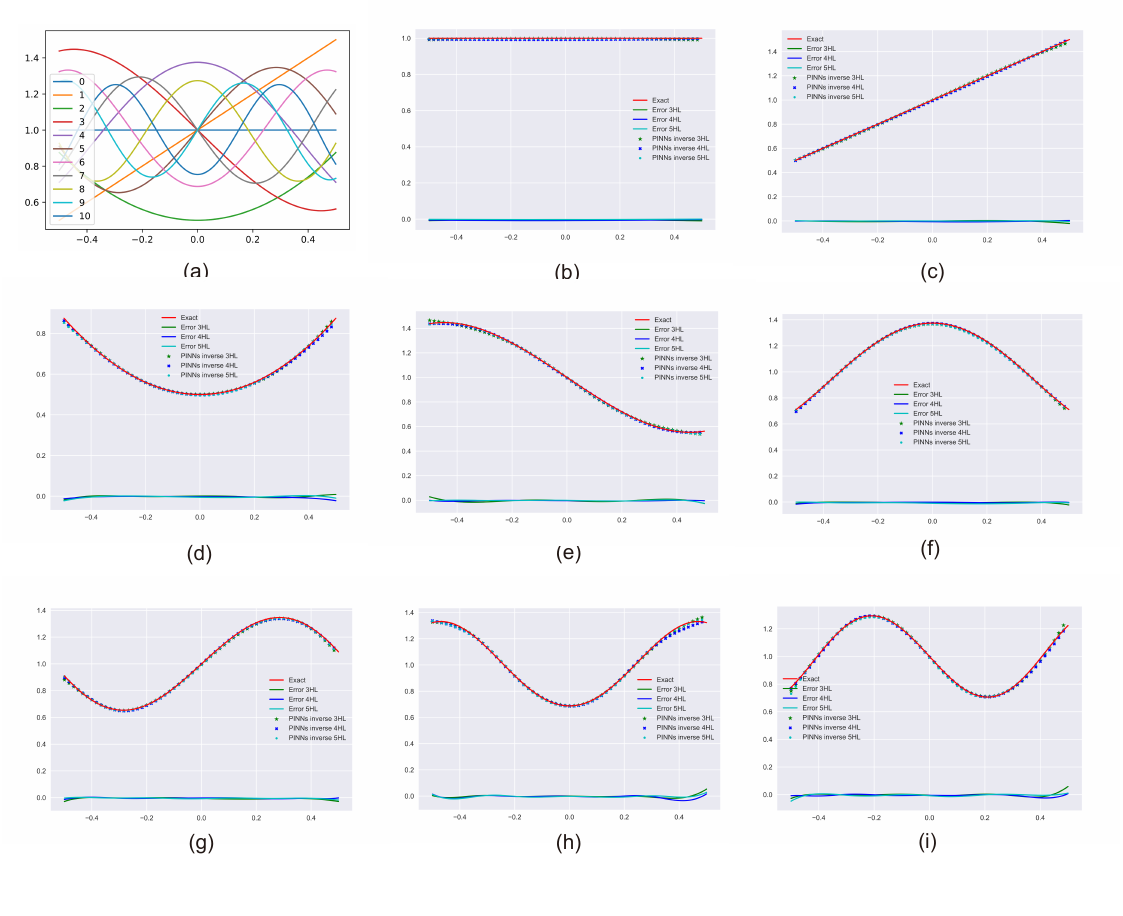}}
		\par\end{centering}
	\caption{The result of prediction for  non-uniform conductivity by  PINNs-based micro-scale model. (a) Legendre orthogonal polynomials from 0 to 10 order. (b) Order 0; (c) Order 1; (d) Order 2; (e) Order 3; (f) Order 4; (g) Order 5; (h) Order 6; (i) Order 7; 
		\label{fig:k_field_different}}
\end{figure}

\begin{figure}
	\begin{centering}
		\subfloat{\centering{}\includegraphics[scale=1.1]{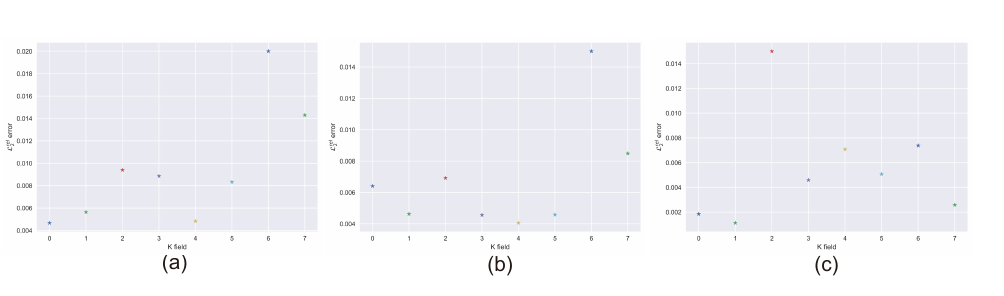}}
		\par\end{centering}
	\caption{The $L_{2}^{rel}$ of non-uniform thermal conductivity (different order of Legendre orthogonal polynomials) based on PINNs-based micro-scale model under different architecture of neural network. (a)
		3 hidden layers; (b) 4 hidden layers; (c) 5 hidden layers.		\label{fig:k_field_L2}}
\end{figure}

To test the performance of the PINNs-based Micro-scale model in the scenario of non-uniform thermal conductivity, we consider the following equation:
\begin{equation}
\begin{cases}
k(x)\frac{d^{2}T}{dx^{2}}=-(15\pi)^{2}cos(15\pi x)k(x) & x\in[-\frac{1}{2},\frac{1}{2}]\\
\frac{\partial T}{\partial x}=-q_{1}=-15\pi & x=-\frac{1}{2}\\
\frac{\partial T}{\partial x}=q_{2}=15\pi & x=\frac{1}{2}
\end{cases}.
\end{equation}
The analytical solution is 
\begin{equation}
T=cos(15\pi x).
\end{equation}
We use Legendre orthogonal polynomials to construct the thermal conductivity field (as shown in the \Cref{fig:k_field_different}a). We choose Legendre orthogonal polynomials because they form a complete set of polynomials that can approximate most realistic spatially varying thermal conductivity fields. We set up 3000 uniformly distributed sensors, and \Cref{fig:k_field_L2} shows the results of all Legendre orthogonal polynomials up to order 7 under different neural network architectures. The results demonstrate that the PINNs-based Micro-scale model can accurately predict the non-uniform conductivity field, with an overall $L_{2}^{rel}$ relative error of approximately $10^{-3}$.

Considering the presence of measurement errors in real-world test data, we also add Gaussian noise perturbations to the temperature field $T(x)$. We set the thermal conductivities to match the real PU-PCMs, i.e., $k=0.24$ (PCMs) and $k=0.03$ (PU). The mathematical formula for the thermal conductivity field is

\begin{equation}
k(x)=0.123+\frac{1}{8}(25x^{4}-15x^{2}+1),
\end{equation}
where it is $k=0.24$ (PCMs) in the middle and $k=0.03$ (PU) at the boundaries. \Cref{tab: k_field_noisy} shows that the PINNs-based Micro-scale model maintains excellent accuracy even under realistic perturbation errors, with errors at the level of approximately $10^{-2}$.

\begin{table}
	\caption{The $L_{1}^{rel}$ error of non-uniform thermal conductivity with noisy data. $\sigma_{s}$ represents the variance of the Gaussian distribution of the noisy data, and HL denotes the different number of hidden layers.}
	\begin{centering}
		\begin{tabular}{cccccccccccc}
			\toprule 
			$\sigma_{s}$ & 0 (no noise) & 0.01 & 0.02 & 0.03 & 0.04 & 0.05 & 0.06 & 0.07 & 0.08 & 0.09 & 0.10\tabularnewline
			\midrule
			3HL & 0.0129 & 0.0166 & 0.0115 & 0.0155 & 0.0139 & 0.0114 & 0.0181 & 0.0176 & 0.0132 & 0.0108 & 0.0146\tabularnewline
			4HL & 0.0193 & 0.0121 & 0.0116 & 0.0172 & 0.0187 & 0.0151 & 0.0130 & 0.0153 & 0.0155 & 0.0107 & 0.0254\tabularnewline
			5HL & 0.0163 & 0.0135 & 0.00593 & 0.00943 & 0.00656 & 0.00716 & 0.0162 & 0.0291 & 0.00819 & 0.00821 & 0.0129\tabularnewline
			\bottomrule
		\end{tabular}
		\par\end{centering}
	\label{tab: k_field_noisy}
\end{table}

\begin{figure}
	\begin{centering}
		\subfloat{\centering{}\includegraphics[scale=1.1]{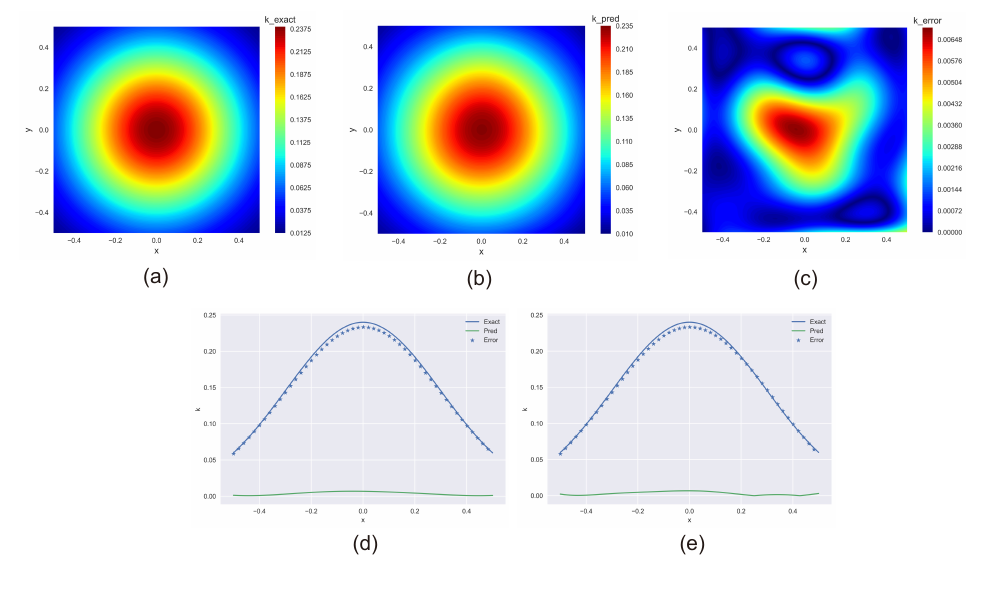}}
		\par\end{centering}
	\caption{The result of micro scale results  based on PINNs in 2D. (a) Exact thermal conductivities; (b) Prediction of thermal conductivities based on PINNs; (c) Absolute error of the PINNs model; (d) Comparision of exact and prediction of thermal conductivities based on PINNs model  in $y=0$; (d) Comparision of exact and prediction of thermal conductivities based on PINNs model  in $x=0$.	\label{fig:k_field_2D}}
\end{figure}

We extend the problem to a two-dimensional case:
\begin{equation}
	\begin{cases}
		k(x,y)(\frac{d^{2}T}{dx^{2}}+\frac{d^{2}T}{dy^{2}})=-[(15\pi x)^{2}+(15\pi y)^{2}]cos(15\pi xy)k(x) & x,y=(-\frac{1}{2},\frac{1}{2})^{2}\\
		\frac{\partial T}{\partial x}=-q_{1}=15\pi ysin(\frac{15\pi}{2}y) & x=-\frac{1}{2},y\in[-\frac{1}{2},\frac{1}{2}]\\
		\frac{\partial T}{\partial x}=q_{2}=-15\pi ysin(\frac{15\pi}{2}y) & x=\frac{1}{2},y\in[-\frac{1}{2},\frac{1}{2}]\\
		\frac{\partial T}{\partial y}=-q_{3}=15\pi xsin(\frac{15\pi}{2}x) & x\in[-\frac{1}{2},\frac{1}{2}],y=-\frac{1}{2}\\
		\frac{\partial T}{\partial y}=q_{4}=-15\pi xsin(\frac{15\pi}{2}x) & x\in[-\frac{1}{2},\frac{1}{2}],y=\frac{1}{2}
	\end{cases}.
\end{equation}
The analytical solution for this problem is given by:
\begin{equation}
	T=cos(15\pi xy).
\end{equation}
To mimic the thermal conductivity of real PU-PCMs material, we adopt the following thermal conductivity formula:
\begin{equation}
	k(x)=0.24exp(-\frac{x^{2}+y^{2}}{0.18}).
\end{equation}
The highest thermal conductivity value in the center is 0.24, gradually decreasing towards the surrounding, with the boundary approaching an average thermal conductivity of 0.03.
\Cref{fig:k_field_2D} illustrates the accurate prediction of micro-scale results based on PINNs in the 2D case, where there are 500 sensor points in each axis. The center point (x=0, y=0) is relatively less accurate due to the highest predicted value in that position.

In summary, the PINNs-based Micro-scale model demonstrates outstanding performance in predicting the inverse problem of thermal conductivity at the multi-scale micro level and holds significant potential for engineering applications where measurement errors exist.

\subsubsection{Meso and Macro scales results}
\label{S:5.1.2}

\begin{figure}[htbp]
	\centering
	\begin{minipage}[t]{0.48\textwidth}
		\centering
		\includegraphics[width=8cm]{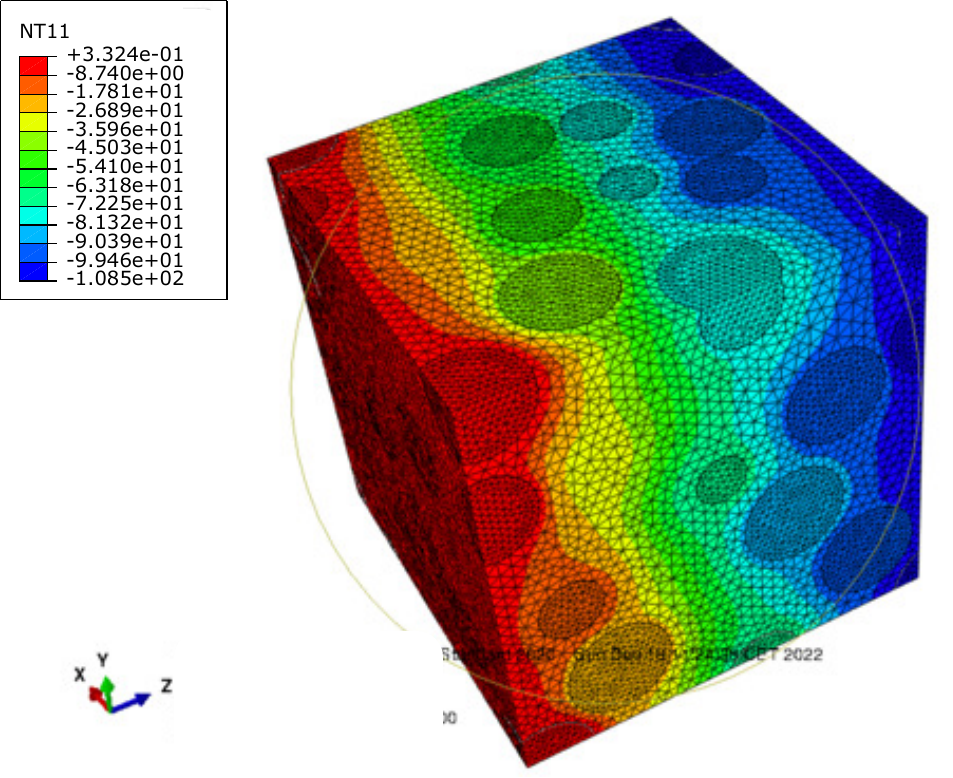}
		\caption{Temperature distribution of composites within RVEs}
		\label{fig: Outputcomposite}
	\end{minipage}
	\begin{minipage}[t]{0.48\textwidth}
		\centering
		\includegraphics[width=8cm]{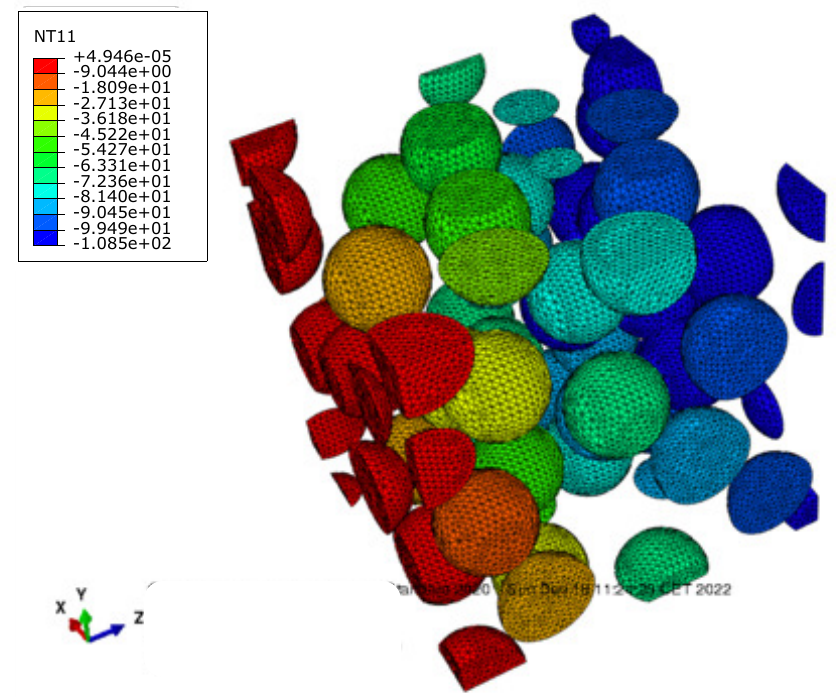}
		\caption{Temperature distribution of  inner plates within RVEs}
		\label{fig: Outputfiber}
	\end{minipage}
\end{figure}

In this part, we focus on presenting the temperature distribution at the mesoscale and the macro-scale engineering parameters derived from FEM-RVE (Finite Element Method-Representative Volume Element). The temperature distribution within the FEM-RVE model is illustrated in \Cref{fig: Outputcomposite} and \Cref{fig: Outputfiber}.

 Furthermore, \Cref{tab:distribuationmaterial} provides information on the composition of cubes within specific material regions, indicating different volume fractions obtained from a finer scale analysis. By combining these fractions, a final thermal conductivity value of 0.2473926 ($W/mK$) is achieved, considering an overall volume fraction of 9\%. These data and results serve as essential engineering parameters for the subsequent scale, facilitating their application in a case study.

\begin{table}[H]
	\centering \caption{The distribution of properties in material region} 
	\label{tab:distribuationmaterial}
	\scalebox{1}{
		\resizebox{\textwidth}{!}{
			\begin{tabular}{ m{2cm}  m{2cm}  m{4cm}   m{2cm}   m{2cm}   m{4cm}  }
				\toprule[2pt]

				Numbers& $V_f$(\%) & Thermal conductivity($W/mk$) &Numbers & $V_f$(\%) & Thermal conductivity($W/mk$)\\
				\midrule[1pt]
				
				1 & 0.043 & 1.8005 & 9 & 0.071 & 0.6038\\
				
				2 & 0.025 & 0.2914 & 10 & 0.045& 1.3574\\
				
				3 & 0.096 & 2.3121 & 11 & 0.019 & 1.2374\\
				
				4& 0.075 & 0.3314 & 12 & 0.041 & 1.6799\\
			   
			    5 & 0.049 & 1.6132 & 13 & 0.048 & 1.2015\\
			    
			    6 & 0.082& 0.6339 & 14 & 0.089 & 1.2315\\
				
				7 & 0.037 & 0.2111 & 15 & 0.088 & 0.8957\\
				
			    8 & 0.057 & 1.1216 & 16 & 0.016 & 1.1912\\

				\midrule[1pt]
				\multicolumn{6}{c} {\centering Thermal conductivity in Mori-Tanaka method: 0.2473926($W/mk$) (Volume fraction: 9\%)} \\

				\bottomrule[1.5pt]
	\end{tabular}}}
\end{table}

\subsection{Case study results}
\label{S:5.2}

In this case study, we investigate the performance evaluation of composites made of PU-PCMs (Phase Change Materials) and their application in single-family houses. Our aim is to incorporate the properties of PU-PCMs into our modeling and assess their impact on energy consumption and thermal comfort. We utilize engineering parameters obtained from a previous multi-scale modeling approach and develop a comprehensive physical simulation of an entire house using partial differential equations in both PINNs and FEM-RVE. All the parameters applied in the computation and simulation are listed in \Cref{tab: ThermMulti}

\begin{table}[htbp]
	\centering \caption{Parameters of PU-PCMs from Multi-scale modeling} \label{tab: ThermMulti}
		\scalebox{1}{
	\resizebox{\textwidth}{!}{
		\begin{tabular}{ m{4cm}  m{6cm}  m{3cm}  m{3cm}  m{3cm} }

			\toprule[2pt]
			Effective scale and length &  Parameters
  &  Value & Method  \\
			\midrule


%
%

		   	Micro ($\mu$m) & Thermal conductivity of inclusion &  0.56W/m·K & Physics-informed neural networks (PINNs)\\
		    
		   	 &  Thermal conductivity of matrix& 0.036W/m·K \\

			Meso ($\mu$m)   &  Interface resistance & 35MW $m^2/K$ & RVE-FEM\\
			

			Macro (mm)  &  Volume fraction & 9\%  & Mori-Tanaka method\\
			
			&  Phase Change point & 24 $^\text{o}$C    & Materials test  \\
			
			Engineering Parameter            &  Effective thermal conductivity & 0.24W/m·K & Simulation  \\
			
			\bottomrule[2pt]
	\end{tabular}} }
\end{table}

To evaluate the effectiveness of PU-PCMs in terms of energy consumption, we employ IDA-ICE to calculate the annual energy usage for the year 2022. We also analyze the hourly indoor temperature fluctuations during the same period. The results, presented in \Cref{tab: AnnualEnergyUsage}, compare the annual energy consumption between two scenarios: one without PU-PCMs interlayer in the building envelope and the other with enhanced PU-PCMs interlayer.

\begin{table}[htbp]
	\centering \caption{Annual Energy usage without/with PU-PCMs} \label{tab: AnnualEnergyUsage}
	\scalebox{0.9}{
		\begin{tabular}{ m{4cm}  m{3cm}  m{3cm} m{3cm} m{3cm} m{2.5cm} }

			\toprule[2pt]
			Components &  \multicolumn{2}{c}{Without PU-PCMs}  &  \multicolumn{2}{c}{With PU-PCMs enhanced}\\
			
			& Purchased energy
(kWh)
 & Energy Usage Intensity (kWh/m2) & Purchased energy
(kWh)& Energy Usage Intensity (kWh/m2) & Improvement \\
			\midrule[1pt]
			Lighting, facility &  32199 &  126.1   & 32199   &126.1   & 0\%\\
			
			Electric cooling  &  16062  & 62.9  &  16322   & 64   & 1.74\%\\

			HVAC aux   &  7258  & 28.4  &  7258   & 28.4   & 0\%\\

			Fuel heating &  15369  & 60.2 &  14973   & 58.6  & 2.64\%\\

			Total, Facility electric &  55519  & 217.4  &  55809   & 218.5   & 0.52\%\\
			
			Total, Facility fuel  &  15369   & 60.2   &  14973   & 58.6   & 2.64\%\\
			
			Total  &  70888     & 277.5    &  70782  & 277.1   & 0.14\%\\
			
			Equipment, tenant  &  24149    & 94.5   &  24149   & 94.5   & 0\%\\
			
			Total, Tenant electric   &  24149   & 94.5  &  24149   & 94.5   & 0\%\\
			
			Grand total     &  95037  & 372.1  &  94931   & 371.7 & 0.11\%\\

			\bottomrule[2pt]
	\end{tabular} }
\end{table}

Our data clearly demonstrate that incorporating PU-PCMs in single-family houses in Umeå, considering the weather conditions of 2022, leads to significant energy savings. We observe a 2.64\% improvement in energy efficiency, corresponding to a reduction in CO2 emissions. These findings highlight the promising potential of PU-PCMs in achieving sustainable and environmentally friendly housing solutions. However, we found that the enhancement of PU-PCMs only affects temperature-related energy consumption, with no improvement in equipment and lighting energy consumption.

\begin{figure}[htbp]
	\centering\includegraphics[width=0.75\linewidth]{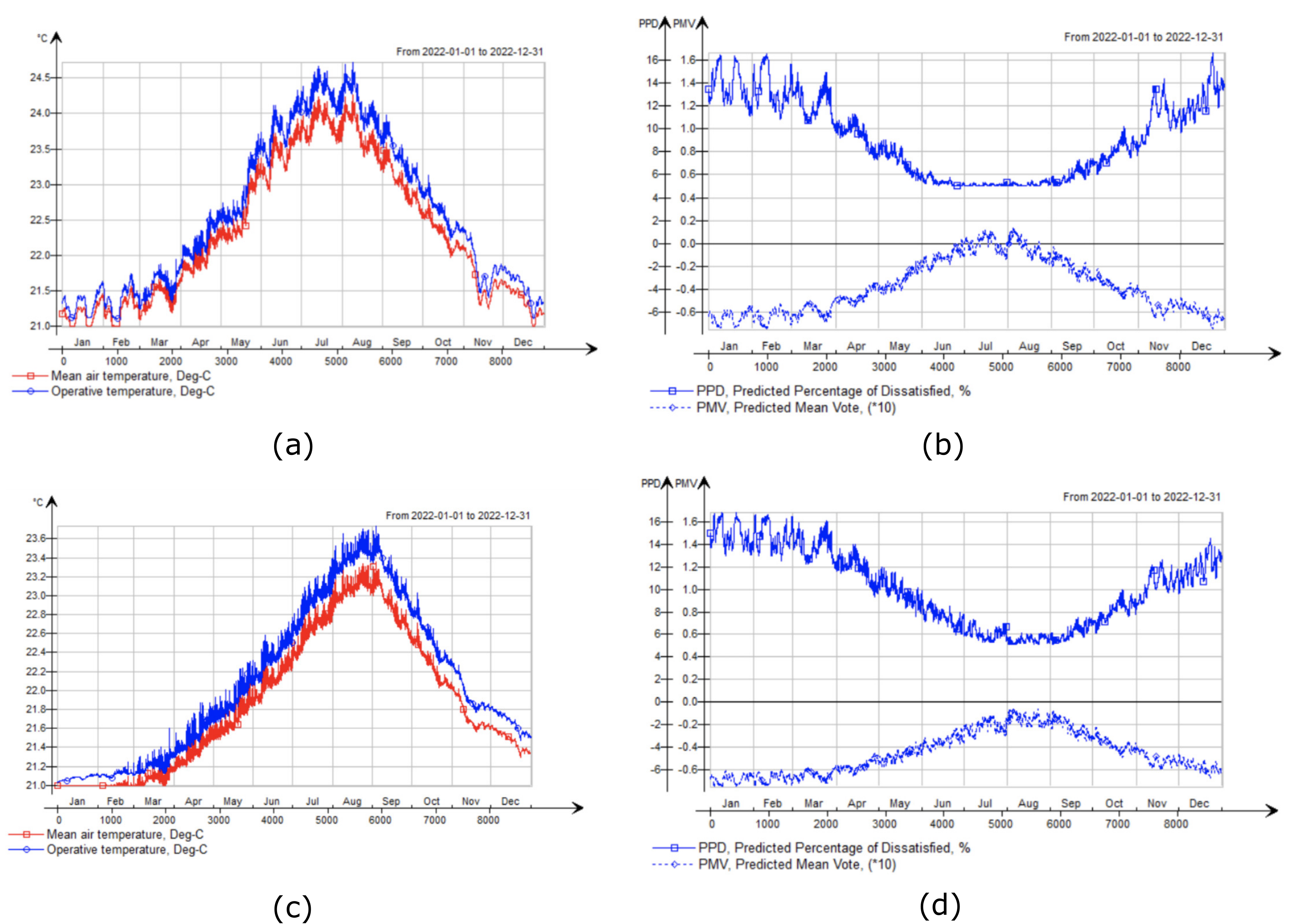}
	\caption{Indoor comfort in Living room:  (a) Main temperature without PU-PCMs. (b) Thermal comfort without PU-PCMs. (c)  Main temperature with PU-PCMs enhanced.  (d) Thermal comfort with PU-PCMs enhanced. }
	\label{fig: Living0PMV}
\end{figure}

\begin{figure}[htbp]
	\centering\includegraphics[width=0.75\linewidth]{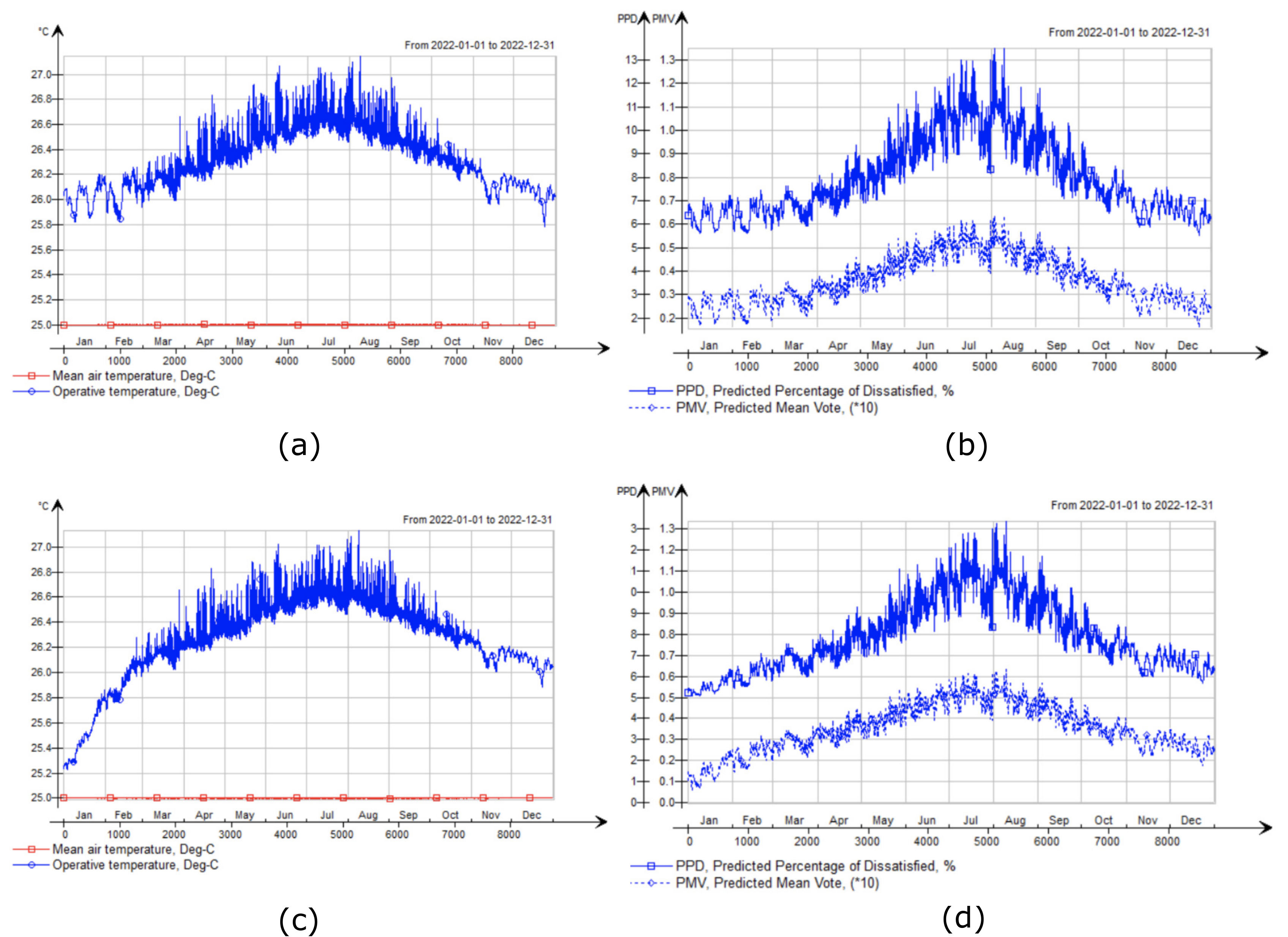}
	\caption{Indoor comfort in Toilet:  (a) Main temperature without PU-PCMs. (b) Thermal comfort without PU-PCMs. (c)  Main temperature with PU-PCMs enhanced.  (d) Thermal comfort with PU-PCMs enhanced. }
	\label{fig: Toilet0PMV}
\end{figure}

\begin{figure}[htbp]
	\centering\includegraphics[width=0.75\linewidth]{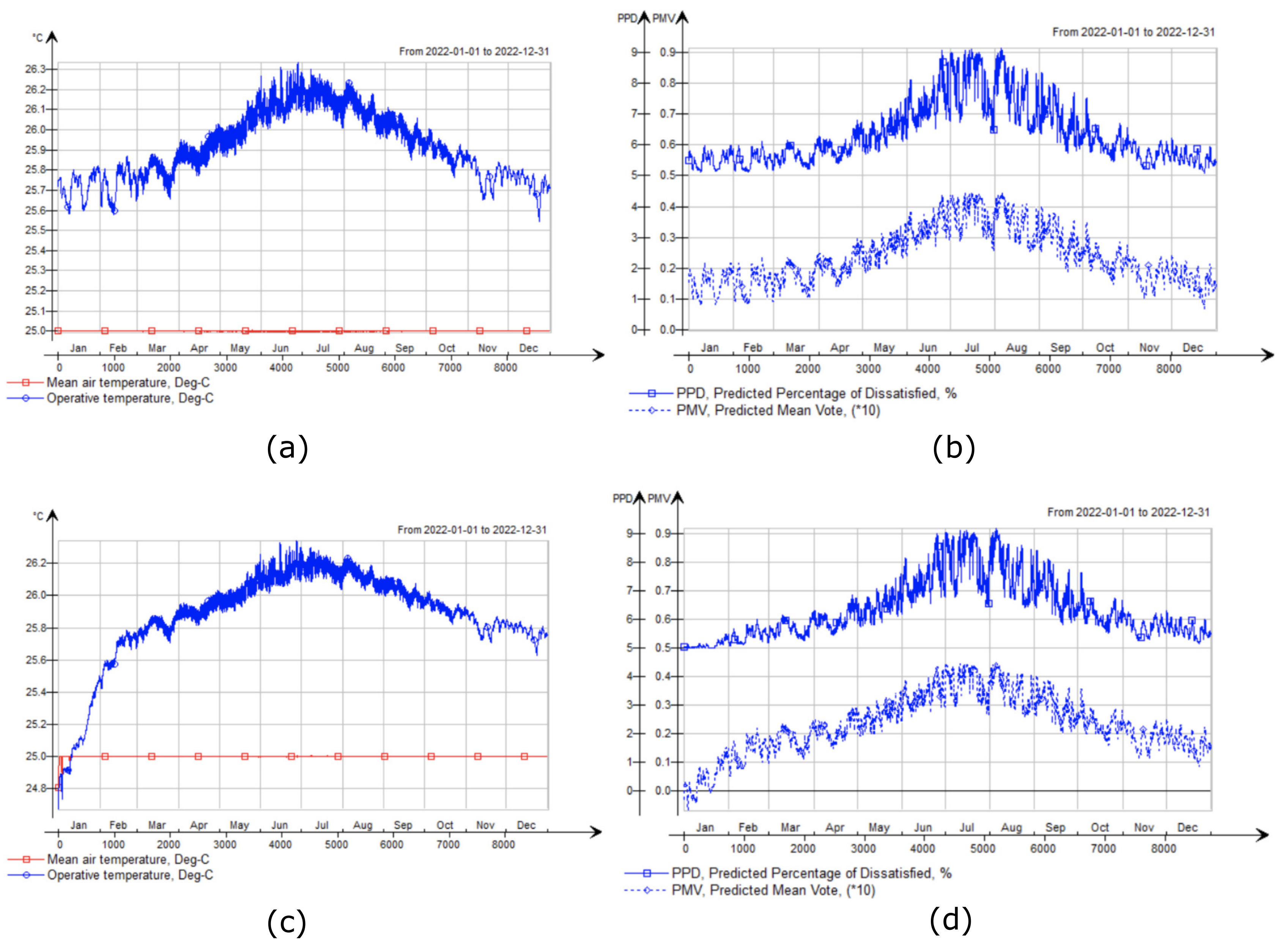}
	\caption{Indoor comfort in 2ed Bedroom:  (a) Main temperature without PU-PCMs. (b) Thermal comfort without PU-PCMs. (c)  Main temperature with PU-PCMs enhanced.  (d) Thermal comfort with PU-PCMs enhanced. }
	\label{fig: 2edbedroomPMV}
\end{figure}

\begin{figure}[htbp]
	\centering\includegraphics[width=0.75\linewidth]{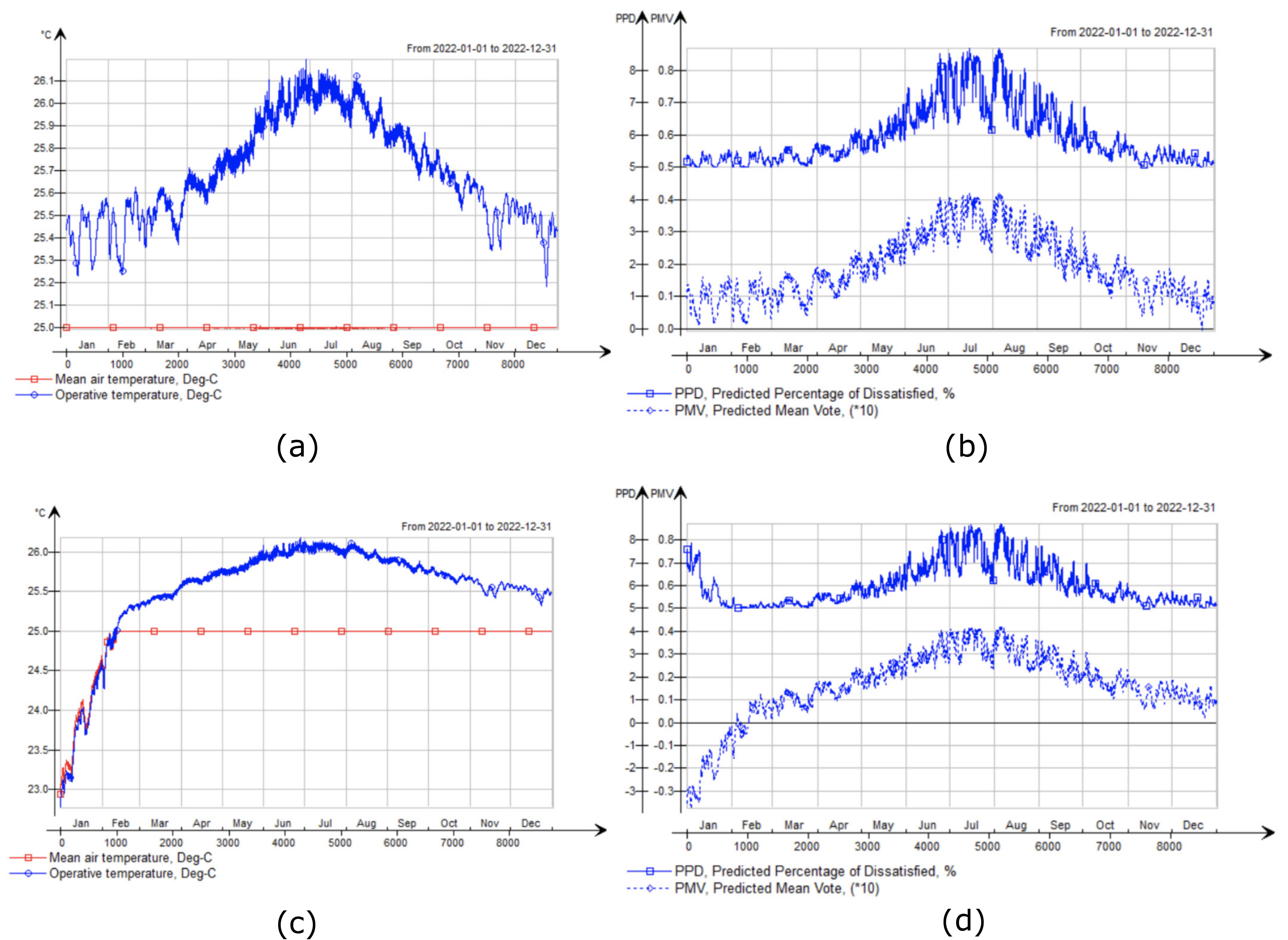}
	\caption{Indoor comfort in 1st Bedroom:  (a) Main temperature without PU-PCMs. (b) Thermal comfort without PU-PCMs. (c)  Main temperature with PU-PCMs enhanced.  (d) Thermal comfort with PU-PCMs enhanced. }
	\label{fig: 1stbedroomPMV}
\end{figure}

We also assess the Fanger comfort index, including PPD and PMV, which indicate occupants' satisfaction with the indoor environment and thermal perception. Figures \Cref{fig: Living0PMV,fig: Toilet0PMV,fig: 2edbedroomPMV,fig: 1stbedroomPMV} display the temperature changes in the Living room, Toilet, 2nd Bedroom, and 1st Bedroom for each month of the year, along with the corresponding occupants' satisfaction and thermal perception under the scenarios without and with enhanced PU-PCMs.

The data summarized in \Cref{tab: FangerComfort} reveal that PU-PCMs significantly improve the living room, reducing dissatisfaction among residents and improving their comfort perception of the thermal environment. The 1st bedroom also benefits from PU-PCMs, with reduced thermal perception and lower heat levels in the living environment. The 2nd bedroom and Toilet show minor enhancements, possibly due to the lower coverage area of PCMs in these rooms during the house design.

\begin{table}[htbp]
	\centering \caption{Fanger's comfort indices without/with PU-PCMs} \label{tab: FangerComfort}
	\scalebox{0.9}{
		\begin{tabular}{ m{2cm}  m{4cm}  m{3.3cm} m{4cm} m{3.3cm} m{2cm} }

			\toprule[2pt]
			Components &  \multicolumn{2}{c}{Without PU-PCMs}  &  \multicolumn{2}{c}{With PU-PCMs enhanced}\\
			
			& Predicted Percentage of Dissatisfied (PPD)
 & Predicted Mean Vote (PMV) & Predicted Percentage of Dissatisfied (PPD)& Predicted Mean Vote (PMV) & Improvement (PMV) \\
			\midrule[1pt]
			Living Room &  9.082\%  &  -3.739   & 5.523\%   & -1.529   & 59.10\% \\
			
			Toilet   &  7.931\%  & 3.615  &  7.829\%   & 3.507   & 3.07\%\\

			2ed Bedroom   &  6.365\%  & 2.405  &  6.354\%   & 2.334   & 3.04\%\\

			1st Bedroom &  6.034\%   & 1.912 &   5.960\% & 1.634  & 17.01\%\\

			\bottomrule[2pt]
	\end{tabular} }
\end{table}

Additionally, we analyze thermal comfort in critical living spaces, including the Living room, Toilet, 2nd Bedroom, and 1st Bedroom within the single-family house. We evaluate the impact of PU-PCMs on real-time temperature fluctuations throughout the year, considering established comfort thresholds.

\begin{figure}[htbp]
	\centering\includegraphics[width=1\linewidth]{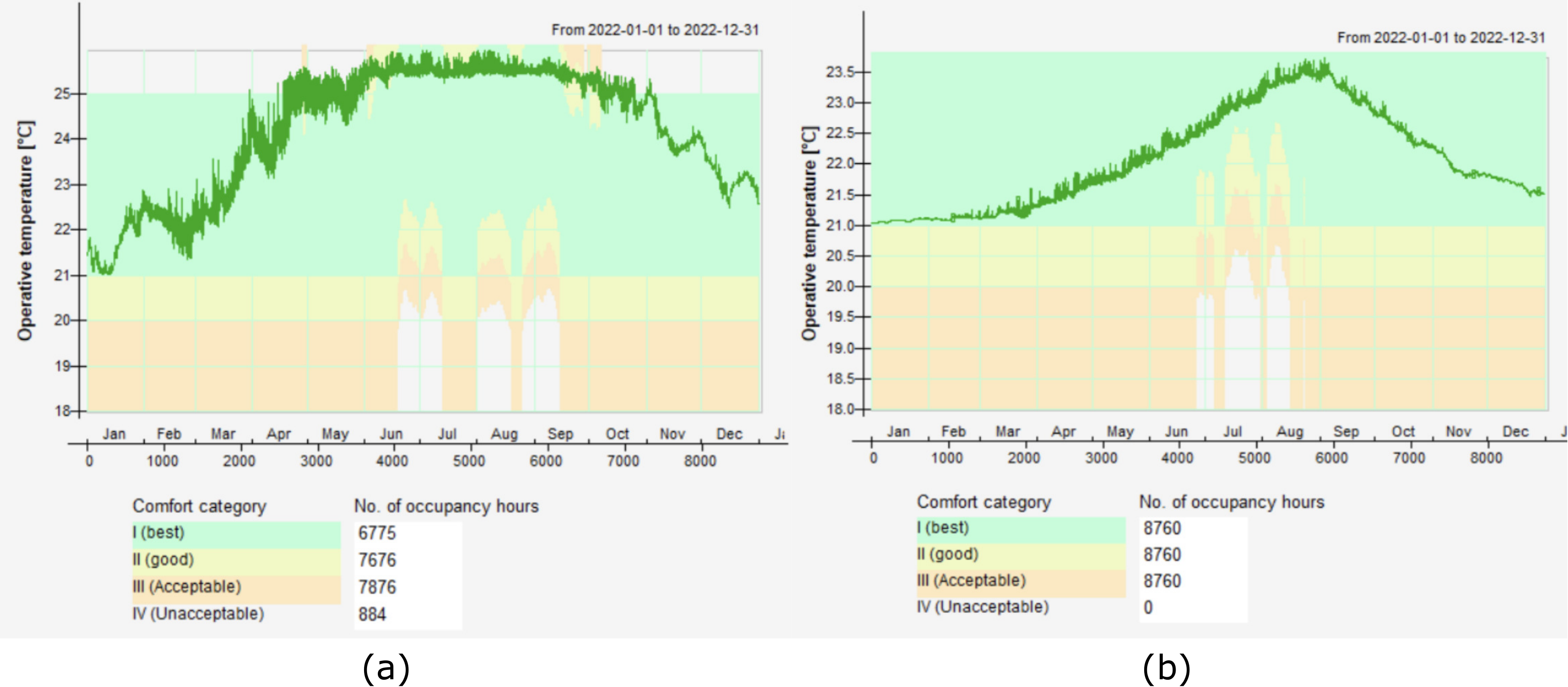}
	\caption{Annul Indoor comfort hours in Living room:  (a) without PU-PCMs. (b) with PU-PCMs enhanced. }
	\label{fig: Living0Hours}
\end{figure}

\begin{figure}[htbp]
	\centering\includegraphics[width=1\linewidth]{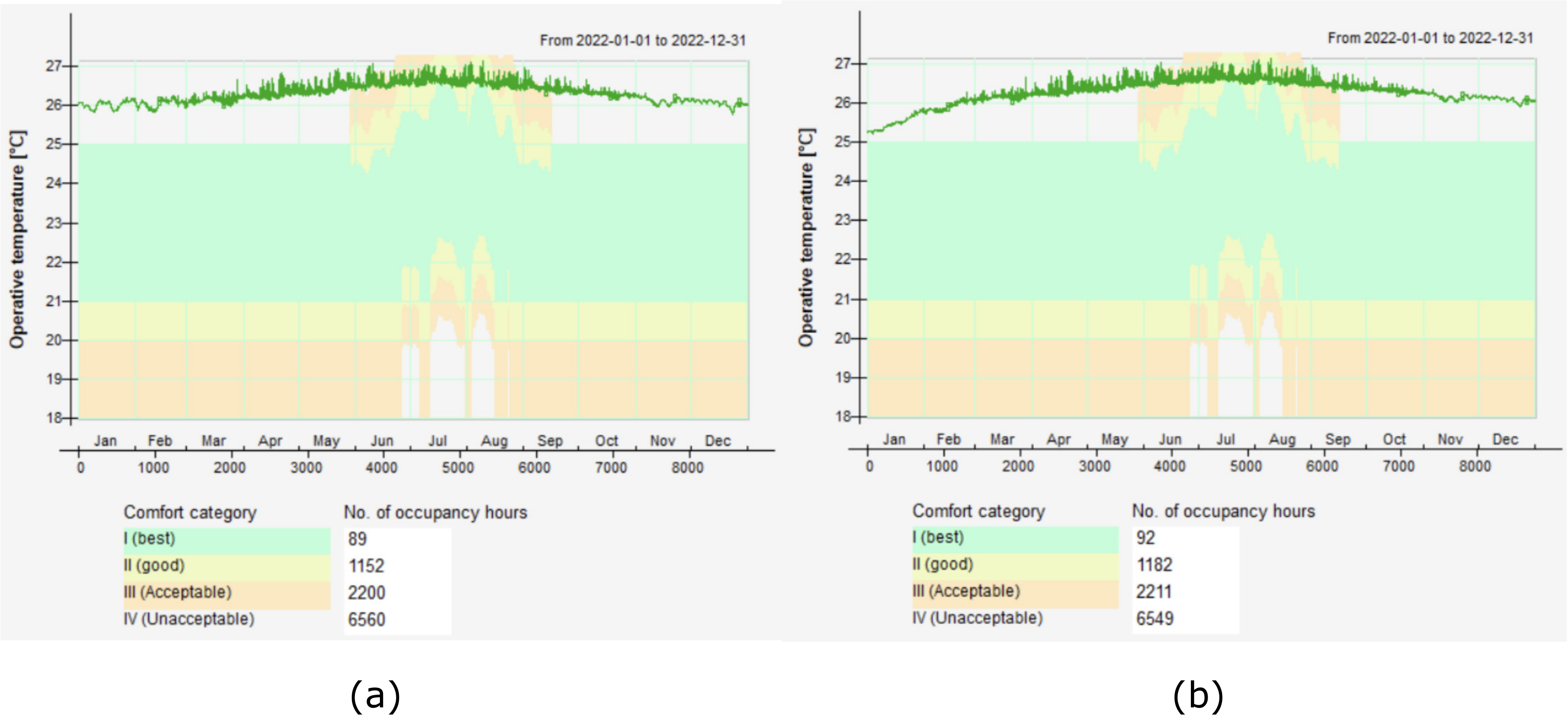}
	\caption{Annul Indoor comfort hours in Toilet:  (a) without PU-PCMs. (b) with PU-PCMs enhanced. }
	\label{fig: Toilet0Hours}
\end{figure}

\begin{figure}[H]
	\centering\includegraphics[width=1\linewidth]{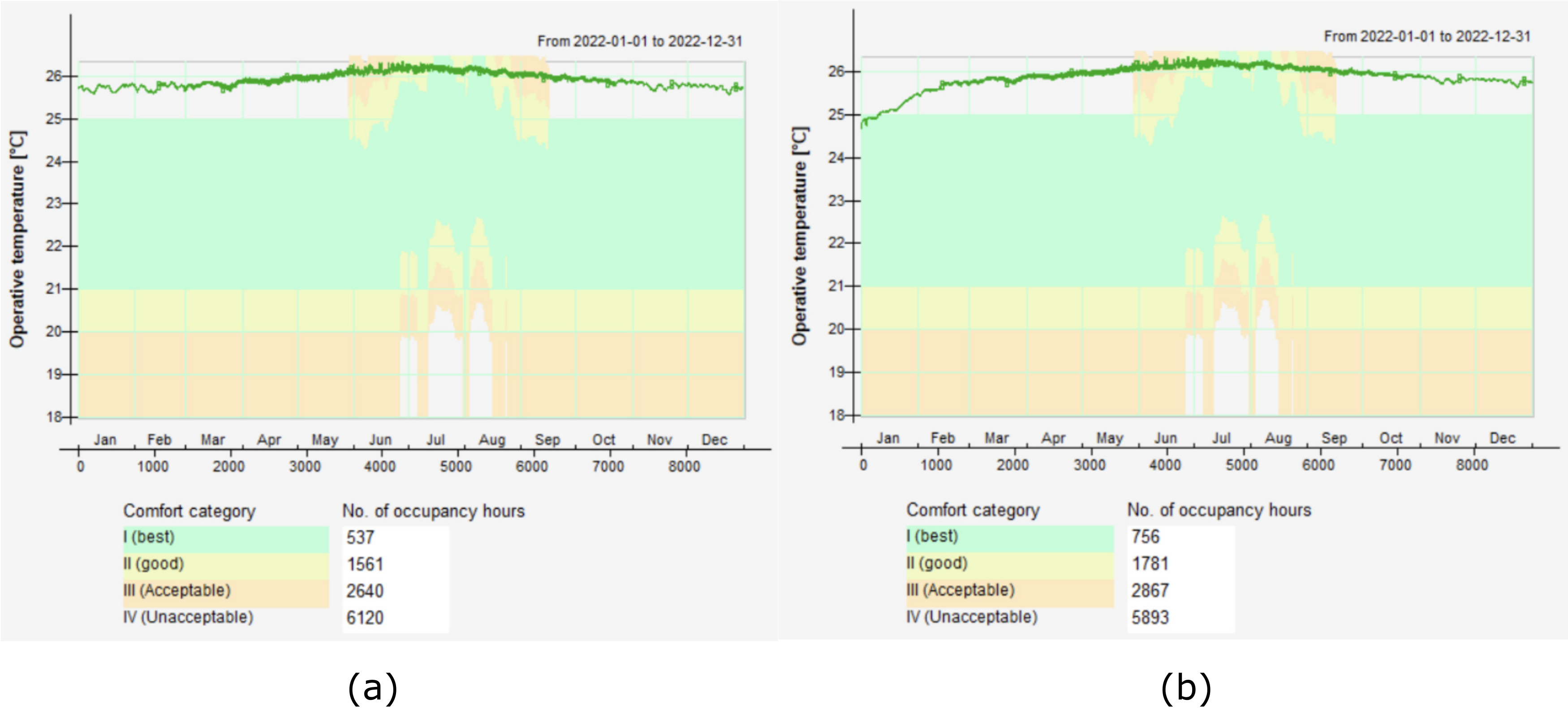}
	\caption{Annul Indoor comfort hours in 2ed Bedroom:  (a) without PU-PCMs. (b) with PU-PCMs enhanced. }
	\label{fig: 2edbedroom1Hours}
\end{figure}

\begin{figure}[htbp]
	\centering\includegraphics[width=1\linewidth]{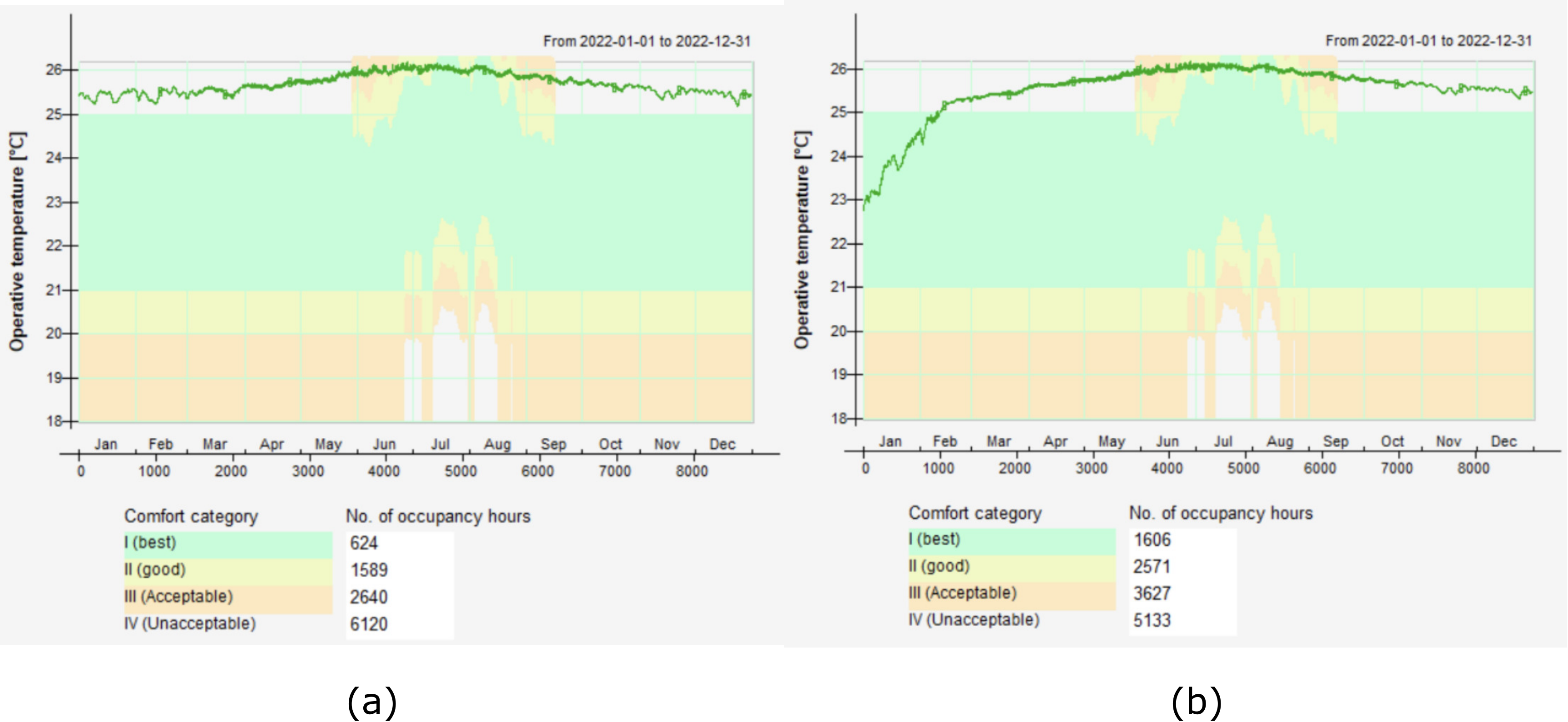}
	\caption{Annul Indoor comfort hours in 1st Bedroom:  (a) without PU-PCMs. (b) with PU-PCMs enhanced. }
	\label{fig: 1stbedroom1Hours}
\end{figure}

Using an operative temperature range of 20-24$^\text{o}$C in winter and 23-26$^\text{o}$C in summer, with fluctuations of 1 degree Celsius considered acceptable, we simulate temperature variations for each of the 8,760 hours in a year. \Cref{fig: Living0Hours,fig: Toilet0Hours,fig: 2edbedroom1Hours,fig: 1stbedroom1Hours} illustrate the results in the Living room, Toilet, 2nd Bedroom, and 1st Bedroom, respectively, demonstrating the impact of incorporating PU-PCMs in the building envelope. Our findings indicate that the addition of a PU-PCMs layer significantly improves thermal comfort hours. Acceptable hours see a 37.38\% increase in the 1st Bedroom, an 11.22\% improvement in the Living room, and an 8.59\% enhancement in the 2nd Bedroom. However, the improvement in the Toilet is only 0.5\% compared to the other rooms. All the results are summarized in \Cref{tab: AnnualComforthours}. These data, along with the accompanying figures, clearly demonstrate that the incorporation of PU-PCMs into the building envelope has a positive impact on thermal comfort throughout the year, especially during the summer months. The enhanced thermal properties of PU-PCMs enable better temperature regulation, resulting in increased comfort and a more pleasant living environment for occupants.

\begin{table}[htbp]
	\centering \caption{Annual Indoor comfort hours without/with PU-PCMs} \label{tab: AnnualComforthours}
	\scalebox{0.9}{
		\begin{tabular}{ m{2cm} m{2cm}  m{2.5cm} m{2.5cm}  m{2cm} m{2.5cm} m{2.5cm} m{2.5cm} }

			\toprule[2pt]
			Components &  \multicolumn{2}{c}{Without PU-PCMs}  &  &\multicolumn{2}{c}{With PU-PCMs enhanced}\\
			
			& Best Hours & Acceptable Hours
 &  Unacceptable Hours & Best Hour & Acceptable Hours& Unacceptable Hours & Improvement (Acceptability) \\
			\midrule[1pt]
			Living Room & 6775 &  7876 &  884  &8760 & 8760   &0   & 11.22\%\\
			
			Toilet  &89  &  2200  & 6560  & 92 &  2211   & 6549   & 0.5\%\\

			2ed Bedroom   &537 &  2640  & 6120 & 756 &  2867   & 5893   & 8.59\%\\

			1st Bedroom & 624&  2640  & 6120 & 1606&  3627   & 5133  & 37.38\%\\

			\bottomrule[2pt]
	\end{tabular} }
\end{table}

Based on a comprehensive analysis of the data, tables, and figures presented above, it is evident that integrating PU-PCMs as an additional layer in the building envelope offers notable benefits. Beyond enhancing energy efficiency, it significantly improves the thermal comfort experienced within the indoor environment. The findings of this case study, conducted in the Umeå region during 2022, highlight the positive impact of PU-PCMs on thermal comfort. The application of these composite materials effectively extends the duration of optimal thermal comfort, particularly during the hot summer period. By incorporating PU-PCMs in the building envelope, the indoor environment experiences more stable and desirable temperature conditions. This not only enhances occupants' comfort but also contributes to a more pleasant and enjoyable living experience.

The results of this study provide strong evidence supporting the use of PU-PCMs in residential buildings, emphasizing their potential to improve both energy efficiency and thermal comfort. The application of PU-PCMs presents a viable solution for addressing the challenges associated with temperature fluctuations and ensuring a comfortable living environment throughout the year.

\section{Conclusions}
\label{sec: 6}

In this study, we propose a hierarchical multi-scale model for a PU-PCM foam composite using Physics-Informed Neural Networks (PINNs). This model enables us to accurately predict and analyze the thermal conductivity of the material at both the meso-scale and macro-scale. By leveraging the integration of physics-based knowledge and data-driven learning provided by PINNs, we effectively solve inverse problems and address complex multi-scale phenomena. The RVE-finite element method is employed to compute the effective engineering parameters of the materials.

The main objective of the multi-scale model is to accurately capture the intricate relationship between the microstructure and thermal properties of PU-PCM. This will be achieved by integrating PINNs, which incorporate fundamental physics and governing equations that govern heat transfer within the material. Through the utilization of available experimental data and simulations, the model will be trained to understand and predict the correlation between microstructural features and the resulting thermal conductivity. The accuracy of the PINNs-based multi-scale model plays a vital role in comprehending and forecasting the thermal behavior of PU-PCM across different scales. Its capability to handle noisy or limited data, along with its capacity to provide uncertainty quantification, will be crucial in addressing real-world scenarios and ensuring reliable predictions.

To assess the practical application of the designed material, we conducted a case study focusing on thermal comfort within a building envelope, specifically in a single house. Additionally, we simulated the annual energy consumption of this scenario. The results demonstrate the promising nature of the design, as it facilitates passive building energy design and significantly improves occupants' comfort.

The successful development of this PINNs-based multi-scale model holds significant potential in advancing our understanding of PU-PCM's thermal properties. It can contribute to the design and optimization of materials for various practical applications, including thermal energy storage systems and insulation in building envelope. In summary, the following conclusion can be summarized:

\begin{enumerate}

\item  \textbf{Novel Approach for multi-scale modeling}: The paper introduces a novel approach, the Physics-Informed Neural Networks (PINNs)-based multi-scale model, for accurately predicting the thermal conductivity of PU-PCM. This approach combines the power of neural networks with physics-based principles, resulting in an accurate and effective solution for the inverse problem.

\item  \textbf{Solving Inverse Problems}: The PINNs-based micro-scale model demonstrates excellent performance in solving inverse problems related to thermal conductivity. It effectively predicts the unknown thermal conductivity field based on given temperature field data, heat source term, and boundary conditions. This capability is crucial for various engineering applications where estimating material properties from observed data is essential.

\item  \textbf{Efficiency and Accuracy}: The PINNs-based approach offers remarkable accuracy in predicting thermal conductivity, even in the presence of measurement errors and noisy data. The results demonstrate high precision, with relative errors maintained at a level of approximately $10^{-3}$ or $10^{-2}$. Furthermore, the computational efficiency of the PINNs-based model is emphasized, as it has the huge potential in multi-scale simulation methods in terms of speed and computational cost.

\item \textbf{Practical Relevance}: The research findings have practical relevance in the field of engineering and materials science. The accurate prediction of thermal conductivity is crucial for designing and optimizing thermal systems, such as energy storage applications using phase change materials. The PINNs-based approach provides a promising tool for enhancing the understanding and prediction of thermal behavior in real-world scenarios.	
	
\item  \textbf{Connecting micro and macro scales}: The RVE-FEM model serves as a reliable tool for effective material design at the mesoscopic scale, allowing for the incorporation of microscopic parameters and their translation into macroscopic engineering structures. In the realm of multi-scale modeling, it plays a crucial role as a bridge that connects the micro and macro levels, facilitating a comprehensive understanding of the system.

\item  \textbf{Enhancing Energy efficiency}: The energy efficiency has been improved by 2.64\% in the case study in the Umeå region in 2022. Meanwhile a corresponding reduction in $CO_2$ emissions can be deducted. 

\item  \textbf{Increasing thermal comfort duration}: The addition of PU-PCMs layer in the building envelope can significantly improve the thermal comfort time, respectively increasing the acceptable thermal comfort time by 37.38\% in 1st Bedroom, 11.22\% improvement of acceptable thermal comfort time in Living Room and 8.58\% improvement of acceptable thermal comfort range in 2ed Bedroom, meanwhile, unacceptable time significantly decreased accordingly.

\item  \textbf{Improving thermal satisfaction}: Regarding occupants' satisfaction with the indoor environment and thermal perception, this design can importantly improve the thermal satisfaction, respectively increase the predicted mean vote (PMV) by 59.10\% in Living Room, 17.01\% improvement of PMV in 1st Bedroom, 3.04\% improvement of PMV range in 2ed Bedroom as well as 3.07\% improvement in Toilet.

\end{enumerate}

~\\

\section*{Acknowledgment}  
We gratefully acknowledge the support of the EU project H2020-AURORAL Grant agreement ID: 101016854 (Architecture for Unified Regional and Open digital ecosystems for Smart Communities and Rural Areas Large scale application) and the Kempe Foundation Sweden (Kempestiftelserna - Stiftelserna J.C. Kempes och Seth M. Kempes minne). This work is also funded and supported by J. Gust. Richert stiftelse, SWECO, Sweden (Grant agreement ID: 2023-00884). 

The computations handling were enabled by resources provided by the Swedish National Infrastructure for Computing (SNIC) and Academic Infrastructure for Supercomputing in Sweden (NAISS) at High-Performance Computing Center North (HPC2N) partially funded by the Swedish Research Council through grant agreement no. 2018-05973 and no. 2022-06725.




\section*{References}
\bibliographystyle{model1-num-names}
\bibliography{Reference.bib}






\end{document}